# Uncertainty quantification of a multi-component Hall thruster model at varying facility pressures.

Thomas A. Marks,[1, a)] Joshua D. Eckels,[1] Gabriel E. Mora,[1] and Alex A. Gorodetsky[1, b)]
*Department of Aerospace Engineering, University of Michigan, Ann Arbor, Michigan, USA*

Bayesian inference is applied to calibrate and quantify prediction uncertainty in a coupled multi-component Hall thruster model. The model consists of cathode, discharge, and plume sub-models and outputs thruster performance metrics, one-dimensional plasma properties, and the angular distribution of the current density in the plume. The simulated thrusters include a magnetically shielded thruster operating on krypton, the H9, and an unshielded thruster operating on xenon, the SPT-100, at pressures between 4.3–43 µTorr-Kr and 1.7–80 µTorr-Xe, respectively. After calibration, the model captures key pressure-related trends, including changes in thrust and upstream shifts in the ion acceleration region. Furthermore, the model exhibits predictive accuracy to within 10% when evaluated on flow rates and pressures not included in the training data, and can predict some performance characteristics across test facilities to within the same range of conditions. Compared to a previous model calibrated on some of the same data [Eckels et al. 2024], the model reduced predictive errors in thrust and discharge current by greater than 50%. An extrapolation to on-orbit performance is performed with an error of 9%, capturing trends in discharge current but not thrust. These findings are discussed in the context of using data for predictive Hall thruster modeling in the presence of facility effects.

## I. INTRODUCTION

Hall thrusters are the most widely-flown type of electric spacecraft propulsion device, but despite their popularity but they remain challenging to model. Predictive models—those which can accurately predict the plasma properties and global performance features of a thruster from geometry and operating conditions alone—are a long-standing goal of the Hall thruster modeling community. Unfortunately, poorly-understood physical effects have to date prevented such efforts from reaching fruition. In addition to the well-known problem of anomalous cross-field electron transport,[1] there are also many subtle and hard-to-model interactions between a thruster and the vacuum facility in which it is tested. These "facility effects" lead to thrusters performing differently in conditions attainable in on-ground facilities than they do in space.[2–5] These effects complicate efforts to correlate ground test data with in-space performance, increasing the expense of thruster development and qualification. In the absence of physics-based models for the aforementioned phenomena, engineering simulations of Hall thrusters rely on phenomenological models with parameters that must be calibrated to match data.[6–9] These modeling choices introduce uncertainty which should be quantified.

To address this challenge, we apply a multidisciplinary modeling approach. We model the thruster-facility system in terms of a series of modular components which interact via a limited set of coupling variables. We then apply Bayesian inference to calibrate the coupled model against data. Once calibrated, we make probabilistic predictions at operating conditions and facilities outside of those on which the model was trained. There are several advantages to this approach. First, the modularity allows new facility effects to be incorporated without changing existing models. Models may likewise be upgraded or replaced as improvements become available. Lastly, our approach places uncertainty quantification in a central role. As calibration is done probabilistically, we obtain not just point estimates but full distributions of model parameters and predictions of key quantities of interest.

In our 2023 paper,[10] we used an previous version of this model to calibrate the SPT-100 thruster across background pressures. While we obtained promising results using a surrogate, predictive accuracy on held-out test data for the model itself was larger than 30%, We take several steps to address the limitations and outcomes of the prior work, and we extend the analysis beyond the SPT-100 and to new operating regimes. Specifically, we make the following changes: **1)** the computational expense of the thruster discharge model has been reduced by greater than 50%, which removes the need of a surrogate model during the UQ analysis, **2)** the Bayesian likelihood has been updated to limit over-confidence in learned model parameters, **3)** the pressure-dependent models for neutral ingestion, acceleration region shift, plume divergence angle, and anomalous electron transport have been revised to better match experimental trends, and to better allocate model parameters.

These changes, particularly the updates to our modeling assumptions, result in a model that agrees with experimental data to within 10% for thrust and discharge current, with similarly low errors observed for the ion velocity, cathode coupling voltage, and plume ion current density. These results improve upon our previous work, in particular with respect to thrust and discharge current where the model test errors were 30 and 53%,

a) Electronic mail: marksta@umich.edu
b) Electronic mail: goroda@umich.edu

respectively.[10] We additionally simulate a magnetically-shielded thruster and demonstrate similar performance using less data.

This paper is organized as follows. First, in Sec. II, we describe the component models, the experimental data, and the methods we apply to calibrate the model. In Sec. III, we demonstrate that our coupled system model exhibits improved prediction accuracy and can extrapolate to a held-out validation dataset. Next in Sec. IV we comment on the applicability of our results to Hall thruster engineering design and consider the limitations of our approach and possible avenues for improvement. Finally, in Sec. V, we summarize our findings.

## II. METHODS

In this section, we describe the models in the coupled framework, the experimental data, the calibration procedure, and the uncertainty quantification approach.

### A. Model

The Hall thruster system model, depicted in schematic form in Fig. 1, comprises an analytic cathode coupling model,[11] a one-dimensional fluid code for the main thruster discharge,[12] and an analytic model for the expansion of the plume into a vacuum chamber.[13] We use these models to predict five quantities of interest (QoIs): thrust, discharge current, cathode coupling voltage, axially-resolved 1-D ion velocity, and the plume ion current density.

As illustrated in Fig. 1, each component model influences different QoIs. The cathode model computes the cathode coupling voltage—the voltage needed to extract cathode electrons into the Hall thruster discharge plasma. This voltage determines the effective potential drop applied to the thruster model. The thruster model then outputs the 1-D axial distribution of plasma properties in the thruster discharge channel and near field plume, such as the electron temperature and ion velocity, the discharge current, ion current, and an "uncorrected" thrust. These last two outputs pass to the plume component, which models the angular distribution of the ion current density at multiple distances downstream of the thruster. We compute the divergence efficiency from this distribution and then use this to "correct" the thrust to account for divergence losses.

Each of the component models has a functional dependence on the facility background pressure, enabling the coupled system to capture a wide range of pressure-dependent phenomena. We define the system model as

$$\mathbf{y} = f(\mathbf{x}) = [f_1(\mathbf{x}), f_2(\mathbf{x}), \ldots, f_Q(\mathbf{x})], \tag{1}$$

where $\mathbf{x}$ and $\mathbf{y}$ are vectors containing all model inputs and outputs, respectively and $Q = 5$ is the number of QoIs. We split the model inputs $\mathbf{x}$ into *operating conditions* $\mathbf{d}$ and *model parameters* $\boldsymbol{\theta}$. Operating conditions represent the experimental conditions at which the data were obtained. These may be known to within some inherent, irreducible uncertainty (*aleatoric uncertainty*) due to measurement precision or noise. In contrast, model parameters are unknown a priori, and have uncertainty that can be reduced by calibrating with data (*epistemic uncertainty*).

Both $\boldsymbol{\theta}$ and $\mathbf{d}$ can be further broken down by component, with subscripts $C$, $T$ and $P$ denoting cathode, thruster, and plume, respectively. Tabs. I and II list each component's inputs and outputs. We use Bayesian inference, described further in Sec. II C, to calibrate the system against data and characterize the posterior distribution of the parameters $\boldsymbol{\theta}$. After calibrating, we use a Monte Carlo analysis of model predictions to understand the relative magnitudes of the aleatoric and epistemic uncertainties. This procedure is described further in Sec. II D.

#### 1. Cathode coupling model

The cathode coupling model, developed in Ref. 11, predicts the cathode coupling voltage, $V_{cc}$, as a function of facility background pressure $P_B$ using the following physically-derived relationship:

$$V_{cc} = V_{vac} + T_{ec}\log\left[1 + \frac{P_B}{P_T}\right] - \left[\frac{T_{ec}}{P_T + P^*}\right]P_B, \tag{2}$$

where $V_{vac}$ is the expected coupling voltage at vacuum, $T_{ec}$ is the effective cathode electron temperature, $P^*$ is the pressure at which $V_{cc}$ stops increasing, and $P_T$ is the base pressure. We treat these four quantities as epistemic model parameters.

#### 2. Thruster model

We use the open-source 1-D axial fluid Hall thruster code *HallThruster.jl*[12] to model the thruster discharge. This code models a quasineutral, multi-species plasma of neutrals, ions, and electrons subject to an accelerating potential. HallThruster.jl solves a continuity equation for the neutrals, both continuity and momentum equations for ions, and an electron energy equation. It then models electrons as an inertialess fluid and computes the electrostatic potential and electron current density using charge conservation and the generalized Ohm's law / quasineutral drift diffusion approximation.[7,14] We consider only singly-charged ions in the present work, although the code supports up to three. HallThruster.jl uses $V_{cc}$ calculated by cathode coupling model as the electric potential at the right (cathode) boundary. The model outputs thrust and ion current, which are passed to the plume model, as well as discharge current and many spatially-resolved plasma properties, including the axial ion velocity.

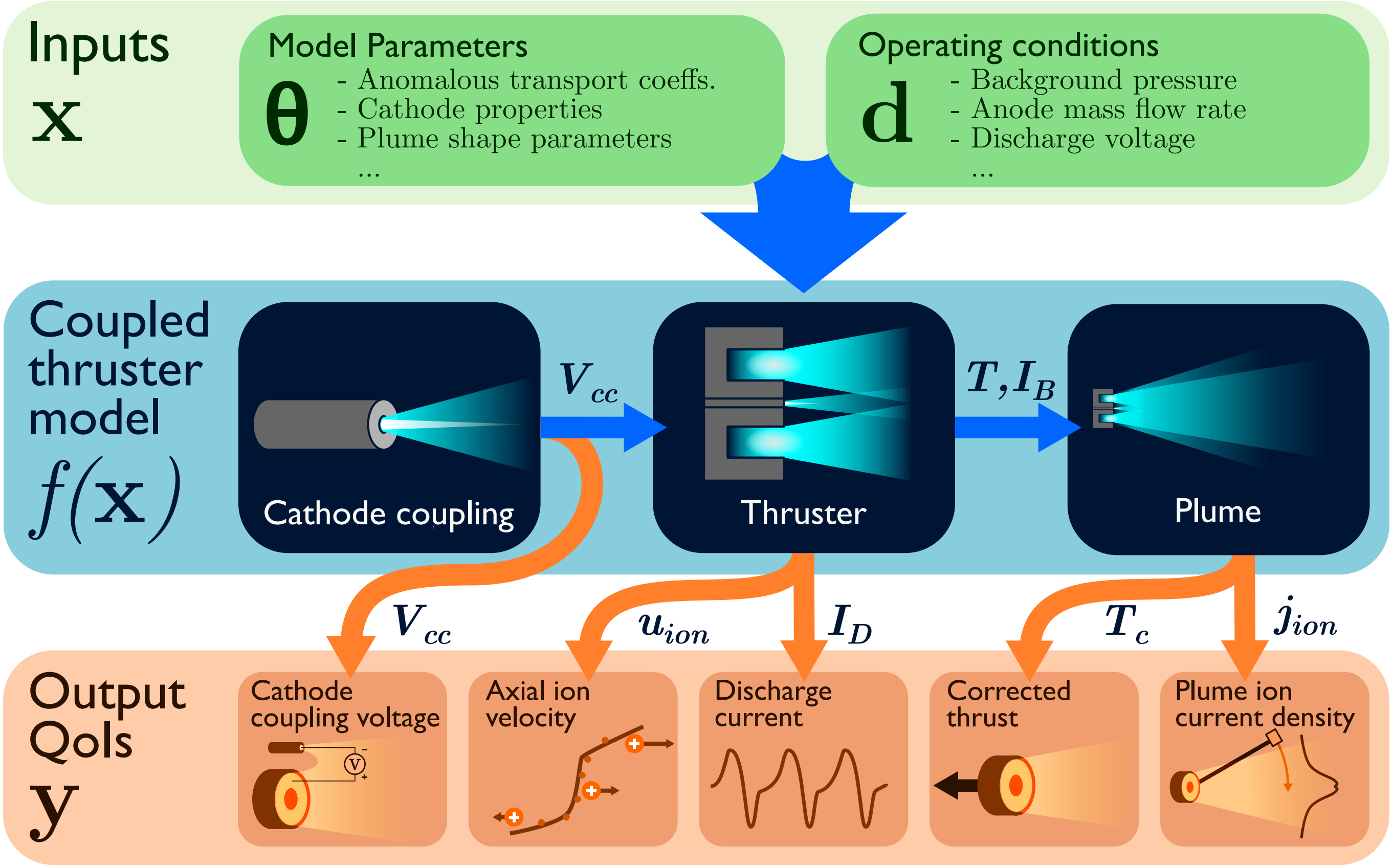

FIG. 1: Overview of the coupled Hall thruster system model, showing the connection between inputs, component models, and output QoIs. Input and output variables are defined in Tab. I and Tab. II, respectively.

HallThruster.jl cannot self-consistently resolve instability-induced cross-field electron transport. Instead, the user specifies a spatially-varying profile for the anomalous electron collision frequency. We employ a four-parameter model of the following form:

$$\Omega_{anom}^{-1} = \alpha_{anom}\left(1 - \beta_{anom}\exp\left[-\left(\frac{\hat{z} - z_{anom}}{L_{anom}}\right)^2\right]\right) \tag{3}$$

where $\Omega_{anom} = \omega_{ce}/\nu_{anom}$ is the anomalous electron Hall parameter, $\nu_{anom}$ is the anomalous electron collision frequency, $\omega_{ce}$ is the electron cyclotron frequency, and $\hat{z}$ is the axial coordinate normalized by the discharge channel length. The transport obeys the Bohm scaling ($\nu_{anom} \sim \omega_{ce}$) with a localized reduction in transport following a Gaussian profile at a specified location. This form captures key features seen in calibrated profiles used in other codes[9] while keeping the number of parameters low. The reduction in transport increases the peak electric field, producing the steep ion acceleration profiles observed in experimental data. The parameters of this model—$\alpha_{anom}, \beta_{anom}, z_{anom}$, and $L_{anom}$—represent the maximum inverse Hall parameter and the scale, location, and width of the transport barrier, respectively. The latter two parameters are non-dimensionalized by the discharge channel length to increase the transportability of parameters between thrusters. We chose this parameterization so that each parameter is of the order $\mathcal{O}(1)$ and to aid interpretability.

As given, this model has no pressure dependence and thus would be unable to capture the observed upstream shift in the ion acceleration region in response to increasing back-pressure[15]. To account for this, we introduce a phenomenological model for this displacement modified from the one in our previous work:[10]

$$\Delta z(P_B) = \Delta z_{anom} L_{ch}\left(\frac{1}{1 + e^{-2(P_B/P_0 - 1)}} - \frac{1}{1 + e^2}\right). \tag{4}$$

Here, $P_B$ is the background pressure and $\Delta z(P_B)$ describes the magnitude of the upstream shift. The parameters of this model are $\Delta z_{anom}$ and $P_0$, which represent the magnitude and center of the shift with respect to background pressure. Eq. (4) is a logistic equation, which captures the intuition that the anomalous transport profile should not move arbitrarily far upstream or downstream as background pressure approaches large or small values. In practice, we have found that $P_0 = 25 \times 10^{-6}$ Torr produces good agreement to data for several thrusters, so we leave only $\Delta z_{anom}$ as a free parameter. This model is implemented by setting $\hat{z} = (z_0 + \Delta z(P_B))/L_{ch}$ in Eq. (3), where $z_0$ is the un-shifted axial coordinate. We show in

TABLE I: Inputs to the coupled thruster-cathode-plume system. The abbreviations C, T, and P refer to the **C**athode, **T**hruster, and **P**lume component models, respectively. Pressures measured in Torr have been corrected for the respective propellant and should be understood as Torr-Xe or Torr-Kr depending on the gas used.

| Symbol | Description | Units | Components | Type | Distribution |
|---|---|---|---|---|---|
| $V_d$ | Discharge voltage | V | C, T | Operating | $\mathcal{N}(\,\cdot\,, 2\%)$ |
| $P_B$ | Background pressure | Torr | C, T, P | Operating | $\mathcal{N}(\,\cdot\,, 5\%)$ |
| $\dot{m}_a$ | Anode mass flow rate | $\mathrm{kg\,s^{-1}}$ | T | Operating | $\mathcal{N}(\,\cdot\,, 2\%)$ |
| $T_{ec}$ | Cathode electron temperature | eV | C, T | Parameter | $\mathcal{U}(1, 6)$ |
| $V_{vac}$ | Vacuum coupling voltage | V | C | Parameter | $\mathcal{U}(0, 60)$ |
| $P_T$ | Base pressure | µTorr | C | Parameter | $\mathcal{U}(20, 200)$ |
| $P^*$ | Turning point pressure | µTorr | C | Parameter | $\mathcal{U}(1, 100)$ |
| $\alpha_{anom}$ | Base inverse Hall parameter | - | T | Parameter | $\mathcal{U}(0, 1)$ |
| $\beta_{anom}$ | Anomalous transport barrier scale | - | T | Parameter | $\mathcal{U}(0, 1)$ |
| $z_{anom}$ | Anom. transport barrier location | - | T | Parameter | $\mathcal{U}(0.75, 1.5)$ |
| $L_{anom}$ | Anom. transport barrier width | - | T | Parameter | $\mathcal{U}(0, 0.5)$ |
| $\Delta z_{anom}$ | Anom. pressure axial shift scale | - | T | Parameter | $\mathcal{U}(0, 0.5)$ |
| $u_n$ | Neutral axial speed | m/s | T | Parameter | $\mathcal{U}(100, 500)$ |
| $c_w$ | Electron wall loss scale | - | T | Parameter | $\mathcal{U}(0.5, 1.5)$ |
| $f_n$ | Neutral ingestion scale | - | T | Parameter | $\mathcal{U}(1, 10)$ |
| $c_0$ | Ratio of main to scattered currents | - | P | Parameter | $\mathcal{U}(0, 1)$ |
| $c_1$ | Ratio of main to scattered div. angles | - | P | Parameter | $\mathcal{U}(0.1, 0.9)$ |
| $c_2$ | Slope of div. angle vs. pressure | $\mathrm{rad\,Pa^{-1}}$ | P | Parameter | $\mathcal{U}(-15, 15)$ |
| $c_3$ | Intercept of div. angle vs. pressure | rad | P | Parameter | $\mathcal{U}(0.2, \pi/2)$ |
| $c_4$ ($10^x$) | Slope of neutral density vs. $P_B$ | $\mathrm{m^{-3}\,Pa^{-1}}$ | P | Parameter | $\mathcal{U}(18, 22)$ |
| $c_5$ ($10^x$) | Intercept of neutral density vs. $P_B$ | $\mathrm{m^{-3}}$ | P | Parameter | $\mathcal{U}(14, 18)$ |

$\mathcal{U}(x, y)$ denotes a uniform distribution between $x$ and $y$.
$\mathcal{N}(\,\cdot\,, x\%)$ denotes a normal distribution about a nominal value with a standard deviation of $x\%$.
Variables with the ($10^x$) notation denote a log-uniform distribution.

TABLE II: Outputs of the coupled cathode-thruster-plume system.

| Symbol | Description | Units | Component | Coupling |
|---|---|---|---|---|
| $V_{cc}$ | Cathode coupling voltage | V | Cathode | Cathode–Thruster |
| $T$ | Uncorrected thrust | N | Thruster | Thruster–Plume |
| $I_B$ | Ion current | A | Thruster | Thruster–Plume |
| $I_D$ | Discharge current | A | Thruster | - |
| $u_{ion}$ | Axial singly-charged ion velocity | m/s | Thruster | - |
| $T_c$ | Corrected thrust | N | Plume | - |
| $j_{ion}$ | Plume ion current density | $\mathrm{A/m^2}$ | Plume | - |

Fig. 2 the shape of the anomalous transport curve and the effect of the pressure shift model.

We calibrate two additional parameters in addition to those controlling anomalous transport: a neutral ingestion scale factor $f_n$ and a wall loss scale factor $c_w$. Hall thrusters typically exhibit increased thrust at high background pressures in part due to the ingestion of background neutrals, which serve as extra propellant.[16,17] We calculate the amount of ingested neutral propellant as the one-sided flux of a stationary Maxwellian population of neutrals of a specified background pressure and temperature across the exit plane of the thruster. On its own, this model may under-predict the true degree of neutral ingestion,[17] so we multiply the ingested neutral flux by $f_n$ to better match experiments. Second, the wall loss scale parameter, $c_w$, scales the plasma edge-to-center density ratio from its base value of 0.5 when performing electron sheath loss calculations. These wall losses are the main method by which HallThruster.jl distinguishes between magnetically-shielded thrusters like the H9 and unshielded thrusters like the SPT-100. In shielded thrusters, we assume the wall temperature equals the anode temperature and we disable ion wall losses, while in unshielded thrusters, the wall temperature equals the channel average temperature and ion wall losses are accounted for.

All simulations in this work use a uniform grid of 100 cells and a domain length of three thruster channel lengths. For the SPT-100 operating at 300 V and 5.0 mg/s, doubling the number of cells or the domain length changes the discharge current by less than 0.1 A and the thrust by less than 5 mN, with the precise and direction of the change depending on the specific simulation parameters used. We simulate one millisecond of thruster operation and average all QoIs over the last 500 microseconds of the simulation. For breathing mode frequencies above 10 kHz, this time is sufficient to allow the discharge to converge to a stationary oscillation or steady state after an initial transient. With these settings, a single HallThruster.jl simulation takes about six seconds on a single core of an Intel Xeon Gold 6154 CPU (the exact time varies per-run due to adaptive time-stepping).

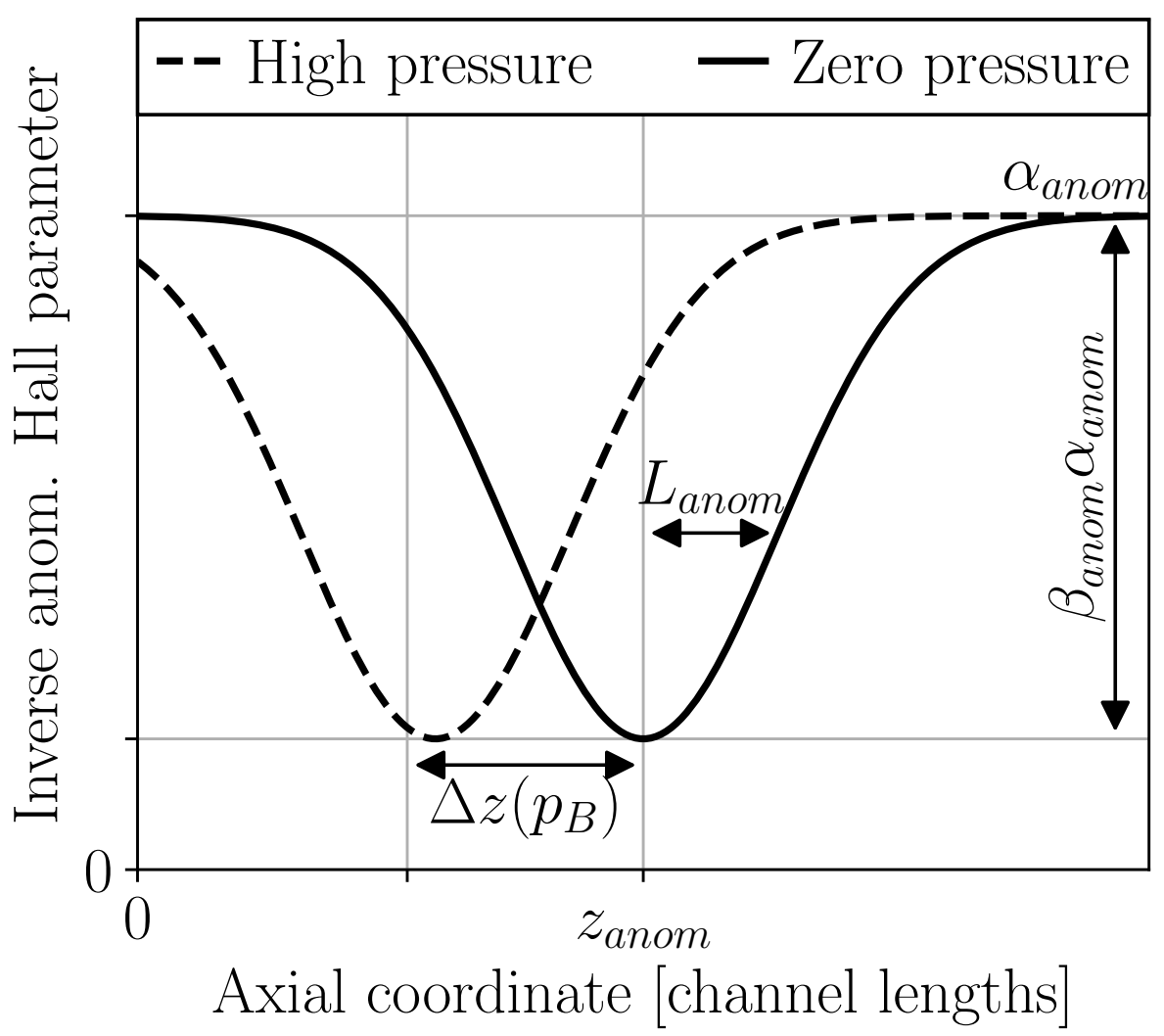

FIG. 2: Notional plot of the anomalous electron transport model used in this work, illustrating the parameters in Eq. (3) and the pressure shift from Eq (4).

To summarize, the changes to the thruster model from our previous work are that the anomalous transport and pressure shift models are more expressive, and we have included additional wall loss and neutral ingestion parameters. In principle, these changes should allow the model to better capture observed trends in facility effects and to better generalize over operating conditions. Additionally, a combination of internal code optimizations and coarsened grid resolution has reduced the runtime by nearly a factor of ten, making it practical to perform Bayesian inference directly on the model without a surrogate. To assess the impact of the modeling changes from the inference changes, we report in Appendix B results obtained by calibrating the model of our previous work using the inference procedure of the present work.

#### *3. Plume model*

We employ the semi-empirical plume expansion model used in Refs. 10 and 13. This model treats the ion current density in the plume ($j_{\text{ion}}$) as composed of three populations—main beam ions, ions scattered by inelastic collisions, and slow ions produced by charge-exchange collisions with neutrals:

$$j_{\text{ion}} = j_{\text{beam}} + j_{\text{scatter}} + j_{\text{cex}}. \tag{5}$$

The first two populations follow Gaussian angular distributions with characteristic divergence angles, while the last expands uniformly in a hemisphere. The current density of each population decays proportionally to the inverse square of the distance from the thruster exit plane.

Given the current density $j_{\text{ion}}(r,\phi)$ as a function of distance $r$ from the thruster exit plane and angle $\phi$ from thruster centerline, we compute the beam divergence angle $\phi_d$ from the ratio of the axial and total ion beam currents:[18]

$$\phi_d = \frac{I_{B,z}}{I_B} = \frac{2\pi r^2 \int_0^{\pi/2} j_{\text{ion}}(r,\phi)\cos(\phi)\sin(\phi)\mathrm{d}\phi}{\int_0^{\pi/2} j_{\text{ion}}(r,\phi)\cos(\phi)\mathrm{d}\phi}. \tag{6}$$

Here, $\phi = 0$ indicates the thruster centerline. We then "correct" the thrust from the thruster model ($T \to T_c$) according to

$$T_c = T\cos(\phi_d), \tag{7}$$

which accounts for the loss in axially-directed thrust due to beam divergence. This differs from our previous work,[10] in which we used the "uncorrected" thrust directly and consequently over-predicted the true measured thrust.

### B. Experimental data

We study two thrusters in this paper — the SPT-100 and the H9. The SPT-100, shown in Fig. 3a, is a widely-tested 1.5 kW-class Hall thruster developed by Fakel in Russia.[16] Due to its age and the availability of its geometry and magnetic field configuration, it is often used for model development activities. The H9 (Fig. 3b) is a magnetically-shielded 9 kW-class Hall thruster developed in collaboration between the University of Michigan (UM), the Air Force Research Laboratory and the Jet Propulsion Laboratory.[19]

We summarize the experimental datasets used in this study for both thrusters in Tab. III, including the measurement QoIs, the number of unique operating conditions (sets of $V_d, P_B, \dot{m}_a$) in each dataset, and the original sources of the data. Datasets categorized as "training" are used in the calibration procedure to tune the model parameters. We additionally include "test" datasets which are not seen during training and use these to assess how well the model generalizes beyond the training data. The SPT-100 datasets[15,16,20] were all obtained using xenon propellant, and all H9 datasets used krypton.[21,22]

The SPT-100 datasets from Ref. 20 include performance measurements (i.e., coupling voltage, thrust, and discharge current) and angularly-resolved measurements of the ion current density in the plume at a radius of one meter from the thruster exit plane, comprising fifteen total operating conditions across two facilities (L3-Harris

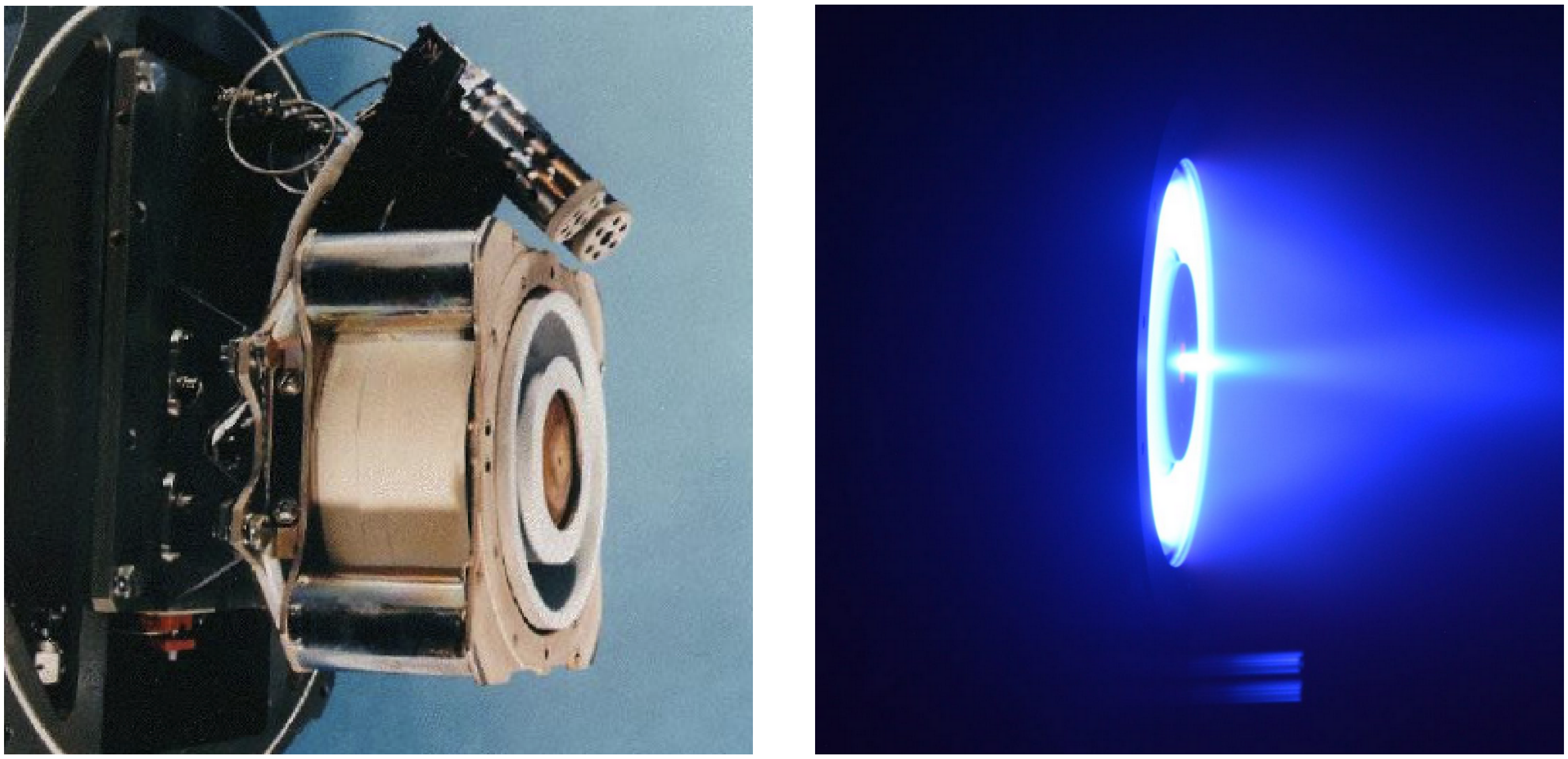

(a) (b)

FIG. 3: (a) The SPT-100 Hall thruster. (b) The H9 Hall thruster operating on krypton in the Large Vacuum Test Facility at the University of Michigan.

TABLE III: Summary of the experimental training and test datasets for the SPT-100 and H9 thrusters in this study. The measurement quantities of interest (QoIs) are listed along with the number of unique operating conditions (sets of $V_d, P_B, \dot{m}_a$) for each dataset. Pressures measured in Torr are corrected for the gas used in the respective dataset.

| Thruster | Gas | Summary | Ref. | QoIs | Conditions | Pressure [µTorr] | Category |
|---|---|---|---|---|---|---|---|
| SPT-100 | Xe | Diamant et al. 2014, L3 | 20 | $V_{cc}$, $I_D$, $T_c$, $j_{ion}$ | 8 | 1.67–55.1 | Training |
| SPT-100 | Xe | Diamant et al. 2014, Aerospace | 20 | $V_{cc}$, $I_D$, $T_c$ | 7 | 3.45–73.7 | Training |
| SPT-100 | Xe | Macdonald-Tenenbaum et al. 2019 | 15 | $I_D$, $u_{ion}$ | 3 | 15.0–50.0 | Training |
| SPT-100 | Xe | Sankovic et al. 1993 | 16 | $I_D$, $T_c$ | 119 | 2.48–55.0 | Test |
| SPT-100 | Xe | Manzella et al. 2001 | 3 | $I_D$, $T_c$ | 2 | 0.02–2.00 | Test |
| H9 | Kr | UM 2024, Plume | 22 | $V_{cc}$, $I_D$, $j_{ion}$ | 3 | 4.48–43.4 | Training |
| H9 | Kr | UM 2024, Velocity | 21 | $I_D$, $u_{ion}$ | 5 | 4.33–30.0 | Training |
| H9 | Kr | GT 2024 | 23 | $I_D$, $T_c$, $j_{ion}$ | 3 | 8.70–22.1 | Test |

Nomenclature for the QoIs is provided in Tab. II.

and Aerospace Corporation). The additional training dataset from Ref. 15 includes the discharge current and spatially-resolved axial ion velocity at three conditions. The test dataset from Ref. 16, includes global performance metrics (thrust and discharge current) at a diverse range of flow rates and background pressures. Finally, the SPT-100 test dataset from Ref. 3 contains measurements from two Russian Express-A satellites, both on the ground and on-orbit. The thruster ran at 300 V on the ground and 310 V on orbit. The mass flow rate for the on-orbit data was not measured and was instead estimated by the authors based on previous SPT-100 experiments. We describe this dataset in more detail in Appendix A.

The H9 data in Tab. III originates from a 2024 test campaign to compare the performance of the same thruster at two different test facilities, namely Vacuum Test Facility 2 at the Georgia Institute of Technology (GT) and the Large Vacuum Test Facility at the University of Michigan. The UM data[21] spans eight background pressures: five of these conditions (the “velocity” dataset in Tab. III) include laser-induced fluorescence measurements of ion velocity[21] and three (the “plume” dataset) contain both cathode coupling voltage and plume ion current density measurements at radii of 1.16 m, 1.32 m, 1.32 m, and 1.64 m.[22] The GT data[23] contains three operating conditions, again differing mainly in background pressure, and includes thrust, ion current density measurements at a distance of one meter, and discharge current. For all H9 datasets, the thruster operated at a nominal discharge current of 15 A and a discharge voltage of 300 V.

### C. Calibration procedure

We calibrate the epistemic model parameters $\boldsymbol{\theta}$ against experimental data using Markov Chain Monte Carlo (MCMC) to generate samples of the epistemic parameters according to their posterior distribution. This is given by Bayes' rule:

$$p(\boldsymbol{\theta} \mid \mathbf{y}_e) = \frac{1}{Z} p(\mathbf{y}_e \mid \boldsymbol{\theta}) p(\boldsymbol{\theta}),$$

where $\mathbf{y}_e$ is a vector of all experimental data at all operating conditions, $p(\boldsymbol{\theta} \mid \mathbf{y}_e)$ is the posterior distribution of all the epistemic parameters given the data, $p(\mathbf{y}_e \mid \boldsymbol{\theta})$ is the likelihood of the experimental data given the parameters and the model, $p(\boldsymbol{\theta})$ is the prior distribution, representing the state of knowledge about the model parameters prior to observing any data, and $Z$ is a normalizing constant.

The prior distributions for each parameter are listed in Tab. I, where we have used uninformative uniform distributions over the expected ranges for each parameter and we assume all parameters are independent. For simplicity, we neglect the aleatoric uncertainty in the operating conditions $\boldsymbol{d}$ during calibration, i.e., we calibrate the epistemic model parameters $\boldsymbol{\theta}$ assuming the operating conditions take on their mean values in Tab. I. Treating the aleatoric uncertainty robustly during calibration would require us marginalize over the aleatoric variables at each MCMC sampling step. This can be done, for instance, using pseudo-marginal MCMC,[24] but would require several model evaluations per sample, significantly increasing the cost of inference. We justify this choice by noting that the aleatoric variables are distributed according to narrow Gaussian distributions, for which a single point sample at the mean provides a good (albeit potentially biased) approximation. In the following section, we discuss our approach to quantifying the impact of aleatoric uncertainty in our predictions; for now, we remark that our method likely underestimates it.

To obtain the likelihood of data given the epistemic parameters, we assume the measurement noise of each observation of each quantity of interest is independent. For each QoI, $n_q$ represents the number of operating conditions which have data for that QoI, and $m_q$ is the length of the observation of that QoI. For the cathode coupling voltage, discharge current, and thrust, $m_q = 1$ since we only observe a single number, but for ion velocity and ion current density $m_q$ will be $\mathcal{O}(10)$–$\mathcal{O}(100)$ as these QoIs are spatially-resolved. Under this formulation, the likelihood is written as

$$p(\mathbf{y}_e \mid \boldsymbol{\theta}) = \prod_{q=1}^{Q} p(\mathbf{y}_{eq} \mid \boldsymbol{\theta}) = \prod_{q=1}^{Q} \prod_{j=1}^{n_q m_q} p(y_{eq}^{(j)} \mid \boldsymbol{\theta}),$$

where $y_{eq}^{(j)}$ represents the $j$-th observation of the $q$-th QoI and $Q = 5$ is the number of QoIs. The outer product assumes independence across quantities of interest, and the inner product assumes independent observations within and across an operating conditions. We next assume that the error can be modeled using additive Gaussian noise. Next, we need to assume some form of error model between the observed model and the predicted value. Specifically, we model the predicted observation of the $q$-th QoI at the $j$-th operating condition as

$$y_{eq}^{(j)} = f_q(\boldsymbol{\theta}, \mathbf{d}_{eq}^{(j)}) + \xi_q, \quad \xi_q \sim \mathcal{N}(0, \sigma_q^2).$$

where $\mathbf{d}_{eq}^{(j)}$ represents the operating conditions associated with a specific experimental data point, $\xi_q$ represents a stochastic model for the error between the model prediction and the observation, and $\sigma_q^2$ is the variance of this error, which may be chosen per-QoI. Under this model we have

$$p(y_{eq}^{(j)} \mid \boldsymbol{\theta}) = \mathcal{N}\left(f_q(\boldsymbol{\theta}, \mathbf{d}_{eq}^{(j)}), \sigma_q^2\right).$$

The log-likelihood is then

$$\begin{aligned}
&\log p(\mathbf{y}_e \mid \boldsymbol{\theta}) \\
&= \sum_{q=1}^{Q} \sum_{j=1}^{n_q m_q} -\log\left(\sqrt{2\pi}\sigma_q\right) - \frac{\left(y_{eq}^{(j)} - f_q(\boldsymbol{\theta}, \mathbf{d}_{eq}^{(j)})\right)^2}{2\sigma_q^2} \\
&= \sum_{q=1}^{Q} \left[ -n_q m_q \log\left(\sqrt{2\pi}\sigma_q\right) - \sum_{j=1}^{n_q m_q} \frac{\left(y_{eq}^{(j)} - f_q(\boldsymbol{\theta}, \mathbf{d}_{eq}^{(j)})\right)^2}{2\sigma_q^2} \right].
\end{aligned}$$

As only the second term depends on $\boldsymbol{\theta}$, we can fold the summation over the first term into a constant $C$ which drops out during MCMC sampling or optimization, giving

$$\log p(\mathbf{y}_e \mid \boldsymbol{\theta}) = -\frac{1}{2} \sum_{q=1}^{Q} \frac{1}{\sigma_q^2} \sum_{j=1}^{n_q m_q} \left(y_{eq}^{(j)} - f_q(\boldsymbol{\theta}, \mathbf{d}_{eq}^{(j)})\right)^2 + C.$$

As written, this likelihood over-weights QoIs like ion velocity which have tens of points per operating condition and underweight global properties like thrust. To mitigate this, we express the pointwise standard deviation $\sigma_q$ in terms of an average relative error across a whole dataset:

$$\sigma_q^2 = c_q \gamma_q^2,$$

where $\gamma_q$ represents a relative measurement error and $c_q$ is a reference magnitude. We use the data to set $c_q$ based on the squared $L_2$ norm of the vector of observations of $q$, averaged over the number of operating conditions in which $q$ was observed:

$$\sigma_q^2 = \frac{||\mathbf{y}_{eq}^{(j)}||^2}{n_q} \gamma_q^2,$$

where the $||\mathbf{y}_{eq}||^2 = \sum_j^{m_q n_q} (y_{eq}^{(j)})^2$. Using this choice, the log likelihood becomes

$$\log p(\mathbf{y}_e \mid \boldsymbol{\theta}) = -\frac{1}{2} \sum_{q=1}^{Q} \frac{n_q}{\gamma_q^2} \frac{||\mathbf{y}_{eq} - f_q(\boldsymbol{\theta}, \mathbf{d}_{eq})||^2}{||\mathbf{y}_{eq}||^2} + C. \quad (8)$$

The interpretation of $\gamma_q/\sqrt{n_q}$ is as an averaged relative $L_2$ norm of the difference between the data and the model output in QoI $q$ across all operating conditions. In the results that follow, we choose $\gamma_q/\sqrt{n_q} = 2.5\%$ for all QoIs, which is equal to or lower than to the estimated relative measurement error for all quantities except the cathode coupling voltage and thrust of the SPT-100, for which we instead set $\gamma_q/\sqrt{n_q} = 1\%$.

With this likelihood in hand, we employ MCMC (specifically the Delayed Rejection Adaptive Metropolis algorithm[25]) to draw 50,000 samples from the posterior distribution. We discard the first half of the drawn samples as burn-in and perform our analysis using the second half. We calibrate each thruster separately, giving each its own posterior parameter distribution. As this likelihood differs somewhat from that used in our previous paper[10]we assess the impact of this change independent of differences in our modeling assumptions and parameterizations in Appendix B. There, we repeat the main analysis of this work using the model of the previous work in concert with the new likelihood, which can be compared with the main results obtained in the following sections.

### D. Uncertainty quantification

After obtaining samples from the posterior parameter distribution, we then quantify the uncertainty in model predictions when comparing to experimental data. We wish to characterize the impact of both epistemic and aleatoric uncertainty in our predictions: the epistemic uncertainty is obtained by propagating only samples of model parameters $\boldsymbol{\theta}$ through the model, and total uncertainty (aleatoric + epistemic) is obtained by sampling the aleatoric uncertainties in the operating conditions $\boldsymbol{d}$. As discussed previously, this approach likely underestimates the aleatoric uncertainty but improves our method's computational efficiency. In Sec. IV B we discuss the effects of this choice in further detail.

Concretely, we adopt the following procedure. For epistemic uncertainty, we draw $N$ samples from the posterior parameter distribution $p(\boldsymbol{\theta} \mid \mathbf{y}_e)$ obtained by MCMC (we take $N = 1000$ throughout our analysis), keeping the aleatoric variables at their nominal values. For the total uncertainty, we draw $N$ samples from both $p(\boldsymbol{\theta} \mid \mathbf{y}_e)$ and $p(\mathbf{d})$ (the prior distribution of the aleatoric variables, as given in Tab. I). In both cases, we then evaluate the model at each input vector $\mathbf{x} = (\boldsymbol{\theta}, \mathbf{d})$ and compute statistics (such as the mean and variance) on the model outputs. Unless otherwise noted, we present in our results the median prediction of each QoI as well as a 90% credible interval of predictions drawn from both distributions. The difference between the epistemic uncertainty bands and the total uncertainty bands gives an estimate of the aleatoric uncertainty compared to epistemic uncertainty.

## III. RESULTS

In this section, we first examine the posterior parameter distributions obtained by the Bayesian calibration procedure before showing the performance on training datasets. We then validate the model's generalization on the independent test datasets and attempt a preliminary extrapolation of the SPT-100 data to orbit. Next, we analyze the anomalous transport curves inferred by the model and conclude by assessing the sensitivity of the model to the calibration parameters.

### A. Calibration

In this section, we present the results of the Bayesian inference procedure described in Sec. II C. We list for the SPT-100 and H9, respectively in Tabs. IV and V, each calibration parameter, their prior distributions, and several statistics of their posterior (post-calibration) distributions.

We find that the posterior distributions of most parameters are narrowed from the prior distributions, indicating that the data is informative for reducing the epistemic uncertainty while not being so restrictive as to produce point estimates. There are a few parameters, however, whose posterior distributions span nearly the same range as their priors:

1. **The wall loss scale parameter ($c_w$) for the H9:** The wall loss model's behavior for shielded thrusters described in Sec. II A 2 reduces the wall interactions to the point that $c_w$ has little effect on the likelihood for this thruster

2. **Cathode parameters (especially the electron temperature, $T_{ec}$, and base pressure, $P_T$) for the SPT-100:** The closeness of the cathode data to a flat value of $V_{cc} = V_{vac}$ means we are less successful in reducing the uncertainty in these parameters.

3. **The neutral ingestion scale parameter ($f_n$) for both thrusters;** While this parameter was intended to improve the model's sensitivity to changing background pressure, in practice it is not able to be inferred with precision. Instead, the coupling between thruster and plume models means that increased plume divergence at lower pressure is sufficient to replicate the trends in the thrust data.

Additionally, the inferred distributions of some parameters differed between thrusters. The median value of the anomalous pressure shift parameters, $\Delta z_{anom}$, was nearly twice as high for the SPT-100 than for the H9, accurately reflecting the difference in the magnitude of the upstream acceleration region shift with pressure in the two thrusters' training datasets. Similarly, the anomalous collision frequency scale, $\alpha_{anom}$ was twice as high for the H9 as for the SPT-100, although for both thrusters it

TABLE IV: Statistics of the 1-D marginal posteriors of the SPT-100 parameters.

| | | Posterior | | | | | |
|---|---|---|---|---|---|---|---|
| Variable | Prior | Min | $5^{th}$ pctile | $50^{th}$ pctile | $95^{th}$ pctile | Max | Std dev |
| $P_T$ | $\mathcal{U}(10, 100)$ | 10 | 14.47 | 48.72 | 93.63 | 99.95 | 24.81 |
| $P^*$ | $\mathcal{U}(10, 200)$ | 10.04 | 25.60 | 64.85 | 140.12 | 197.95 | 35.27 |
| $T_e$ | $\mathcal{U}(1, 6)$ | 1 | 1.17 | 2.92 | 4.85 | 6 | 1.18 |
| $V_{vac}$ | $\mathcal{U}(0, 60)$ | 29.84 | 30.87 | 31.75 | 32.40 | 32.88 | 0.46 |
| $\beta_{anom}$ | $\mathcal{U}(0, 1)$ | 0.95 | 0.97 | 0.99 | 1 | 1 | 0.01 |
| $z_{anom}$ | $\mathcal{U}(0.75, 1.5)$ | 1 | 1.06 | 1.14 | 1.20 | 1.26 | 0.04 |
| $\alpha_{anom}$ | $\mathcal{U}(0, 1)$ | 0.02 | 0.04 | 0.06 | 0.09 | 0.10 | 0.01 |
| $\Delta z_{anom}$ | $\mathcal{U}(0, 0.5)$ | 0.08 | 0.20 | 0.33 | 0.45 | 0.5 | 0.08 |
| $L_{anom}$ | $\mathcal{U}(0, 0.5)$ | 0.25 | 0.34 | 0.43 | 0.49 | 0.5 | 0.05 |
| $c_w$ | $\mathcal{U}(0.5, 1.5)$ | 0.5 | 0.51 | 0.67 | 1.25 | 1.50 | 0.23 |
| $f_n$ | $\mathcal{U}(1, 10)$ | 1 | 1.40 | 5.23 | 9.53 | 10 | 2.59 |
| $u_n$ | $\mathcal{U}(100, 500)$ | 157.59 | 195.57 | 278.11 | 378.17 | 448.28 | 55.07 |
| $c_0$ | $\mathcal{U}(0, 1)$ | 0.67 | 0.71 | 0.76 | 0.79 | 0.82 | 0.03 |
| $c_1$ | $\mathcal{U}(0.1, 0.9)$ | 0.26 | 0.29 | 0.32 | 0.36 | 0.41 | 0.02 |
| $c_2$ | $\mathcal{U}(-15, 15)$ | -15 | -14.52 | -12.36 | -8.88 | -5.32 | 1.77 |
| $c_3$ | $\mathcal{U}(0.2, \pi/2)$ | 0.2 | 0.2 | 0.21 | 0.22 | 0.24 | 0.01 |
| $c_4$ $(10^x)$ | $\mathcal{U}(18, 22)$ | 20.02 | 20.15 | 20.33 | 20.45 | 20.55 | 0.10 |
| $c_5$ $(10^x)$ | $\mathcal{U}(14, 18)$ | 14 | 14.03 | 14.33 | 15.48 | 16.88 | 0.50 |

Variables with the $(10^x)$ notation indicate a log-uniform distribution.

TABLE V: Statistics of the 1-D marginal posteriors of the H9 parameters

| | | Posterior | | | | | |
|---|---|---|---|---|---|---|---|
| Variable | Prior | Min | $5^{th}$ pctile | $50^{th}$ pctile | $95^{th}$ pctile | Max | Std dev |
| $P_T$ | $\mathcal{U}(1, 100)$ | 1 | 1.22 | 3.18 | 10.89 | 40.03 | 3.51 |
| $P^*$ | $\mathcal{U}(10, 200)$ | 41.24 | 42.47 | 45.33 | 46.86 | 48.79 | 1.33 |
| $T_e$ | $\mathcal{U}(1, 6)$ | 3.03 | 4.18 | 5.4 | 5.95 | 6 | 0.57 |
| $V_{vac}$ | $\mathcal{U}(0, 60)$ | 17.44 | 18.50 | 21.94 | 26.47 | 30 | 2.49 |
| $\beta_{anom}$ | $\mathcal{U}(0, 1)$ | 0.94 | 0.96 | 0.98 | 0.99 | 1 | 0.01 |
| $z_{anom}$ | $\mathcal{U}(0.75, 1.5)$ | 1 | 1.04 | 1.07 | 1.1 | 1.15 | 0.02 |
| $\alpha_{anom}$ | $\mathcal{U}(0, 1)$ | 0.04 | 0.07 | 0.13 | 0.18 | 0.21 | 0.03 |
| $\Delta z_{anom}$ | $\mathcal{U}(0, 0.5)$ | 0.01 | 0.06 | 0.18 | 0.29 | 0.37 | 0.07 |
| $L_{anom}$ | $\mathcal{U}(0, 0.5)$ | 0.19 | 0.29 | 0.43 | 0.49 | 0.5 | 0.06 |
| $c_w$ | $\mathcal{U}(0.5, 1.5)$ | 0.5 | 0.64 | 1.19 | 1.48 | 1.5 | 0.25 |
| $f_n$ | $\mathcal{U}(1, 10)$ | 1 | 1.37 | 3.85 | 8.87 | 10 | 2.25 |
| $u_n$ | $\mathcal{U}(100, 500)$ | 217.09 | 245.48 | 268.86 | 302.43 | 322.44 | 17.7 |
| $c_0$ | $\mathcal{U}(0, 1)$ | 0.03 | 0.15 | 0.32 | 0.64 | 0.77 | 0.15 |
| $c_1$ | $\mathcal{U}(0.1, 0.9)$ | 0.1 | 0.17 | 0.39 | 0.69 | 0.85 | 0.17 |
| $c_2$ | $\mathcal{U}(-15, 15)$ | -9.65 | -4.81 | 2.71 | 14.51 | 15 | 6.68 |
| $c_3$ | $\mathcal{U}(0.2, \pi/2)$ | 0.23 | 0.26 | 0.32 | 0.35 | 0.37 | 0.02 |
| $c_4$ $(10^x)$ | $\mathcal{U}(18, 22)$ | 18.52 | 19.23 | 20.15 | 20.33 | 20.44 | 0.34 |
| $c_5$ $(10^x)$ | $\mathcal{U}(14, 18)$ | 14 | 14.02 | 14.26 | 14.98 | 15.63 | 0.31 |

Variables with the $(10^x)$ notation indicate a log-uniform distribution.

varied across at least a factor of two. Finally, $c_2$, which determines how the divergence angle trends with background pressure was found to have an opposite sign for the H9 as in the SPT-100. For the SPT-100, it is uniformly negative, indicating a reduction in plume divergence at high pressure. By contrast, for the H9 the median value is positive, leading to the opposite trend. These inferred parameters reflect real trends in the data, and indicate the success of the calibration procedure.

Appendix C shows the single and two-parameter marginals of the posterior parameter distributions for each component model and thruster. In the joint distributions, we observe that some parameters ($z_{anom}$ and $\Delta z_{anom}$, $\alpha_{anom}$ and $L_{anom}$, $c_0$ and $c_3$, and for the H9, $V_{vac}$ and $P_T$) are highly-correlated. These correlations are indicative of the fact that some parameters influence the likelihood in similar ways and may trade off against one another. For instance, both $\alpha_{anom}$ and $L_{anom}$ control the total electron current and thus the discharge current, the former by scaling down the anomalous mobility globally, and the latter by broadening the region of low anomalous mobility. As such, at higher values of $\alpha_{anom}$, the model requires a value of $L_{anom}$ to maintain the same discharge current and thus observe a positive correlation between these variables. Fortunately, each of our parameters are identifiable, as each of marginal distributions display clear maxima despite the occasional parameter correlations.

### 1. SPT-100

In Figs. 4, 5, and 6, we show how the model predictions of cathode coupling voltage, discharge current, and thrust differ under the prior parameter distribution and the calibrated posterior distribution. In these plots, we show the median model output of each QoI and the 90% credible interval (CI). The interval contains both aleatoric and epistemic uncertainties in the prior plots, while in the posterior plots we show both separately. For each QoI, the uncertainty is dramatically reduced under the posterior, and the median prediction moves closer to the experimental data.

The cathode coupling voltage is well-recovered under the posterior, including the non-monotonic trend with increasing background pressure. However, the predicted trend is more subtle than the experimental one and peaks at a different pressure. This contrasts with our previous work, which was able to more tightly reduce the uncertainty in the cathode parameters and thus better capture the trend in the data. This is likely a result of the likelihood used in this work, which prioritized relative error over the entire dataset rather than the sum of point-wise absolute errors as in the previous work. We observed this outcome as well in the calibrated cathode parameters in Tab. IV, where overall relative error is greatly reduced by fine-tuning the $V_{vac}$ parameter, but the $(P_T, P^*, T_{ec})$ parameters that characterize the more subtle trends with pressure contribute less to the likelihood and so were calibrated to a much lesser extent. It is possible that given additional time, the calibration procedure may have fine-tuned these parameters more to better fit the experimental trend. We recover the correct flat trend in discharge current with respect to background pressure in Fig. 5, though the 4.25 A points from Ref. 15 lie slightly outside of the CI. The posterior predictive CI bounds also encompass the experimental thrust and exhibit the correct trend with pressure, i.e., slightly increasing with background pressure. For cathode coupling voltage and discharge current, the epistemic uncertainty is much larger than the aleatoric uncertainty, while for thrust the aleatoric uncertainty is equal to or greater than the epistemic uncertainty. This stems from the fact that the thrust is more directly impacted by voltage and flow rate than the discharge current and cathode coupling voltage.

In Fig. 7, we compare the simulated ion velocity to measurements from Ref. 15. Our model captures the upstream shift in ion acceleration region with increasing background pressure as well as the maximum slope of the ion velocity profile. In the data, the ion velocity profile at $P_B = 35\,\mu\text{Torr}$ actually sits about 1 mm further upstream than that at $P_B = 50\,\mu\text{Torr}$. The authors of the original paper noted that this was unexpected, as in most thrusters the acceleration region shifts monotonically upstream with pressure. As our model also assumes monotonicity, we do not capture this feature of the data. The main discrepancies with data occur upstream of the acceleration region, inside of the discharge channel ($z/L_{ch} < 1$), where the model overestimates the ion velocity. The reason for this overestimate is unclear, but likely has to do with the large ion backflow region (where $u_{ion} < 0$) seen in the data, which is unusual compared to ion velocity measurements on other thrusters. As an example of the level of model uncertainty typical for these predictions, we show in Fig. 7b the uncertainty bounds for the prediction at $35\,\mu\text{Torr}$.

In Fig. 8, we show the plume ion current density profile at a distance of 1 meter from the thruster exit plane, compared to data from Ref. 20. For visual clarity, we only show three representative pressures out of the eight in the dataset. The model agrees with the data well, especially at angles less than 60 degrees. At larger angles, the absolute errors remain low while the relative error increases; this effect is magnified visually by the use of the logarithmic y-axis scale in Fig. 8. The likelihood used during calibration implicitly weights points with larger magnitudes higher than those with lower magnitudes. As very small current densities at large angles do not contribute much to the divergence angle integrals in Eq. 6, this choice prioritizes fitting the parts of the ion current density curve with a direct impact on the observable QoIs.

### 2. H9

In Figs. 9, 10, and 11, we show the prior and posterior predictions of the cathode coupling voltage, discharge current, and thrust from the H9. The model captures the monotonic trend in cathode coupling with pressure, and has reduced the uncertainty in the discharge current to a narrow band around 15 A. In contrast to the SPT-100 results, we predict a decreasing trend in thrust at high background pressures. We return shortly to a discussion of possible reasons for this trend.

In Fig. 12, we plot ion velocity curves for the H9 at three representative pressures. We observe both good agreement between the model and data as well as low prediction uncertainty. In particular, the final exit velocity, the pressure-dependent acceleration region shift, and the steepness of the acceleration profile are all captured accurately.

The current density dataset for the H9 includes measurements at multiple background pressures, each in turn taken at several distances from the thruster. For visual clarity, we first show results at a single distance and multiple pressures, followed by results at multiple distances and

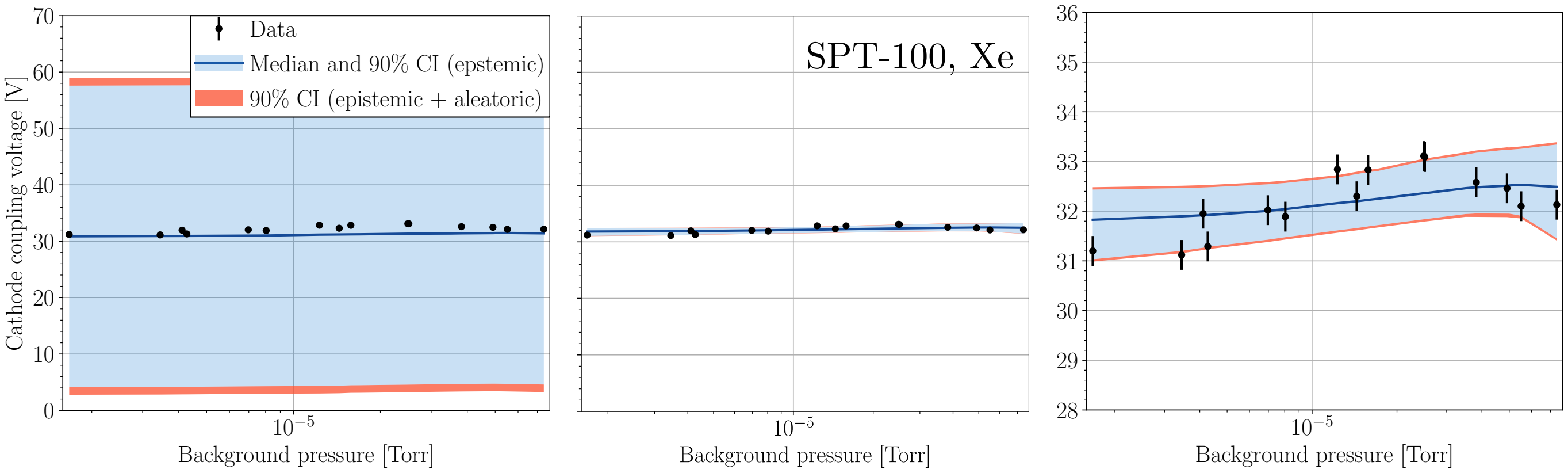

(a) (b) (c)

FIG. 4: (a) Prior, (b) posterior, and (c) zoomed-in posterior predictions of the SPT-100's cathode coupling voltage as a function of background pressure, compared to data from Ref. 20.

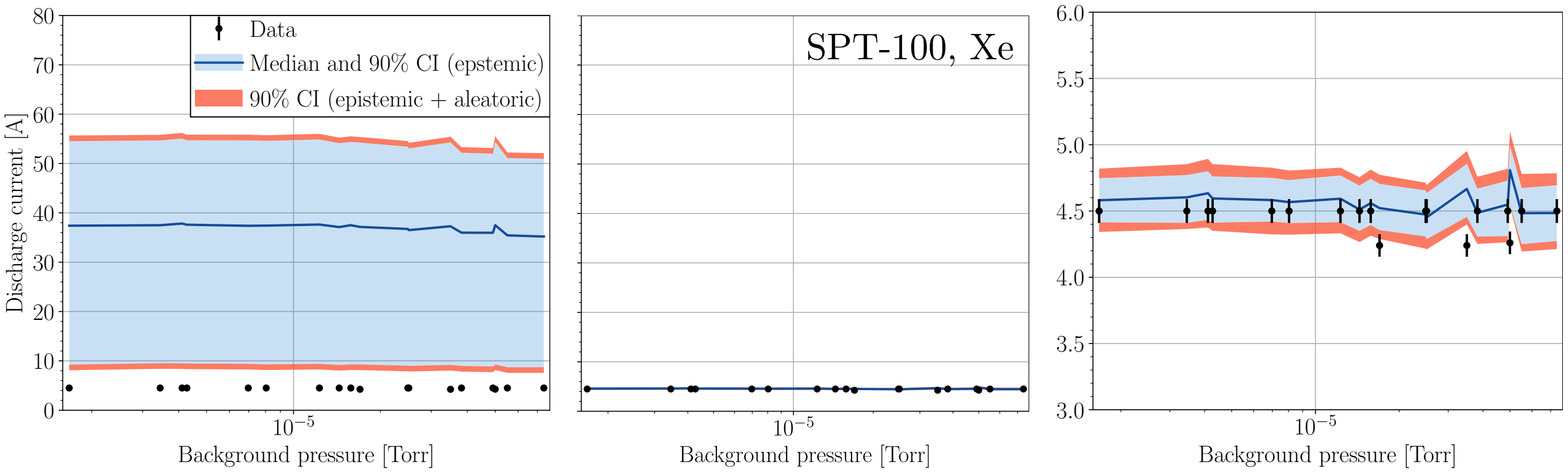

(a) (b) (c)

FIG. 5: (a) Prior, (b) posterior, and (c) zoomed-in posterior predictions of the SPT-100's discharge current as a function of background pressure compared to data from Refs. 20 and 15.

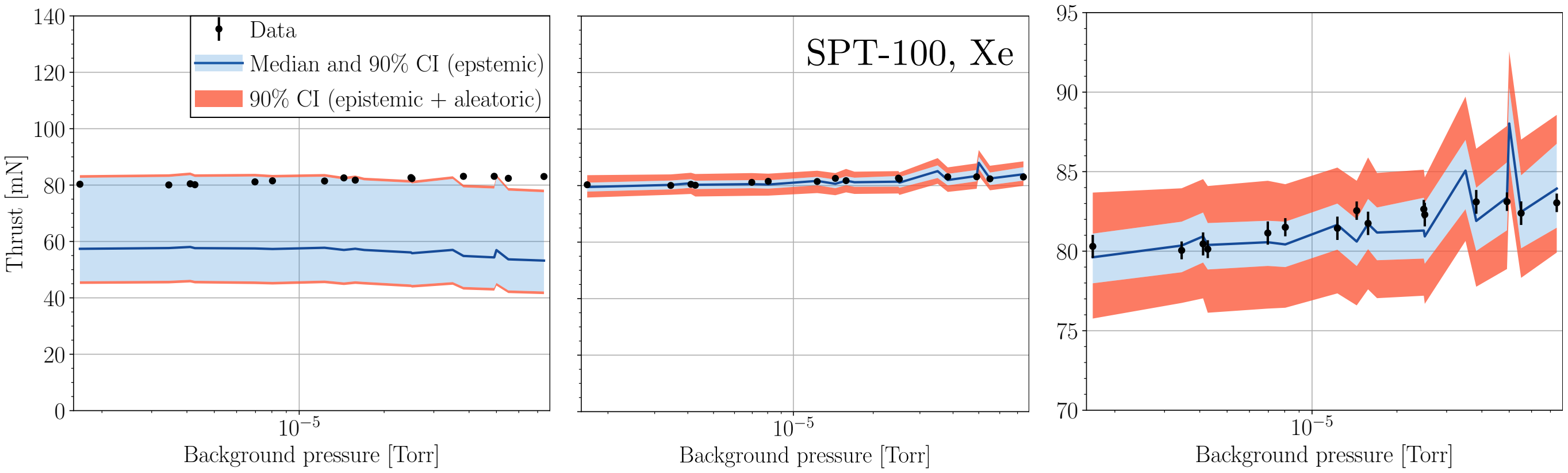

(a) (b) (c)

FIG. 6: (a) Prior, (b) posterior, and (c) zoomed-in posterior predictions of the SPT-100's thrust as a function of background pressure compared to data from Ref. 20.

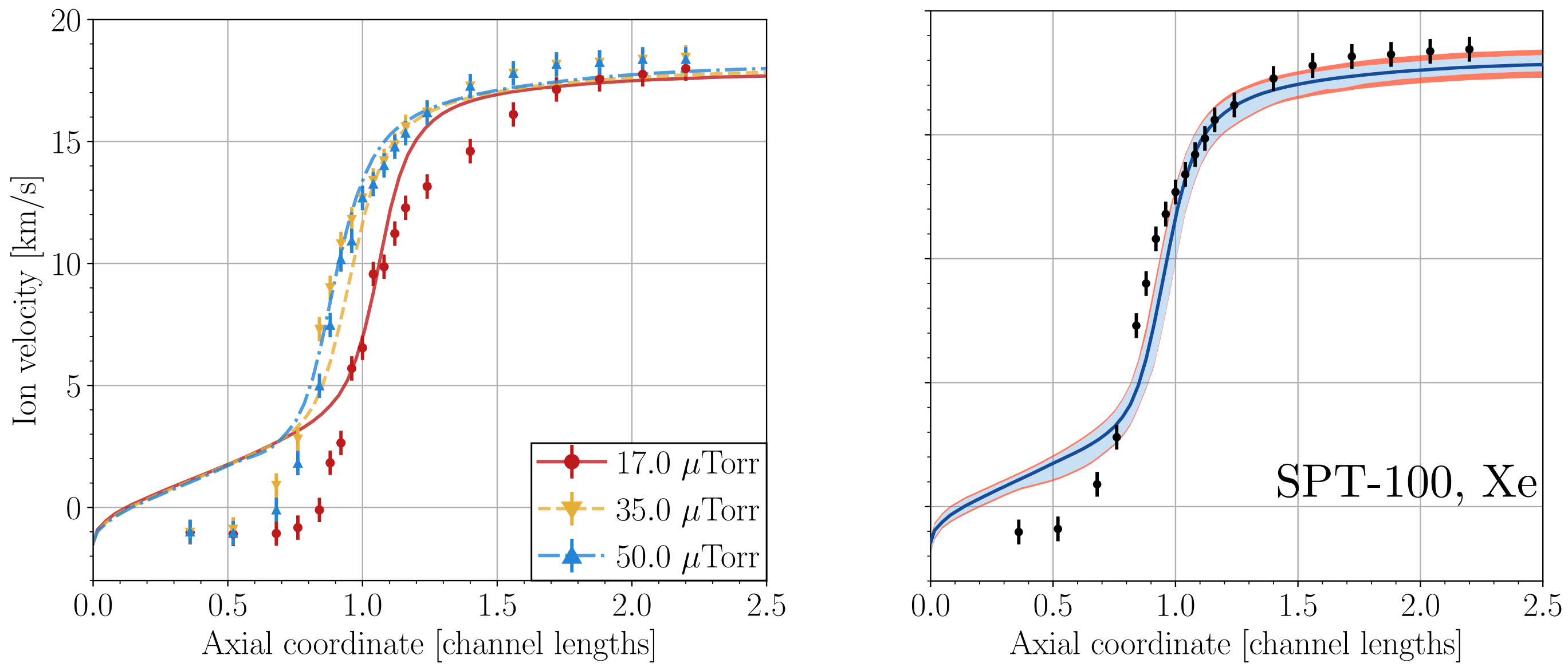

(a) (b)

FIG. 7: (a) Median of posterior predictions of the SPT-100's axial ion velocity at three background pressures, compared to data from Ref. 15, (b) Median posterior prediction and uncertainty in ion velocity at $P_B = 35.0\,\mu$Torr.

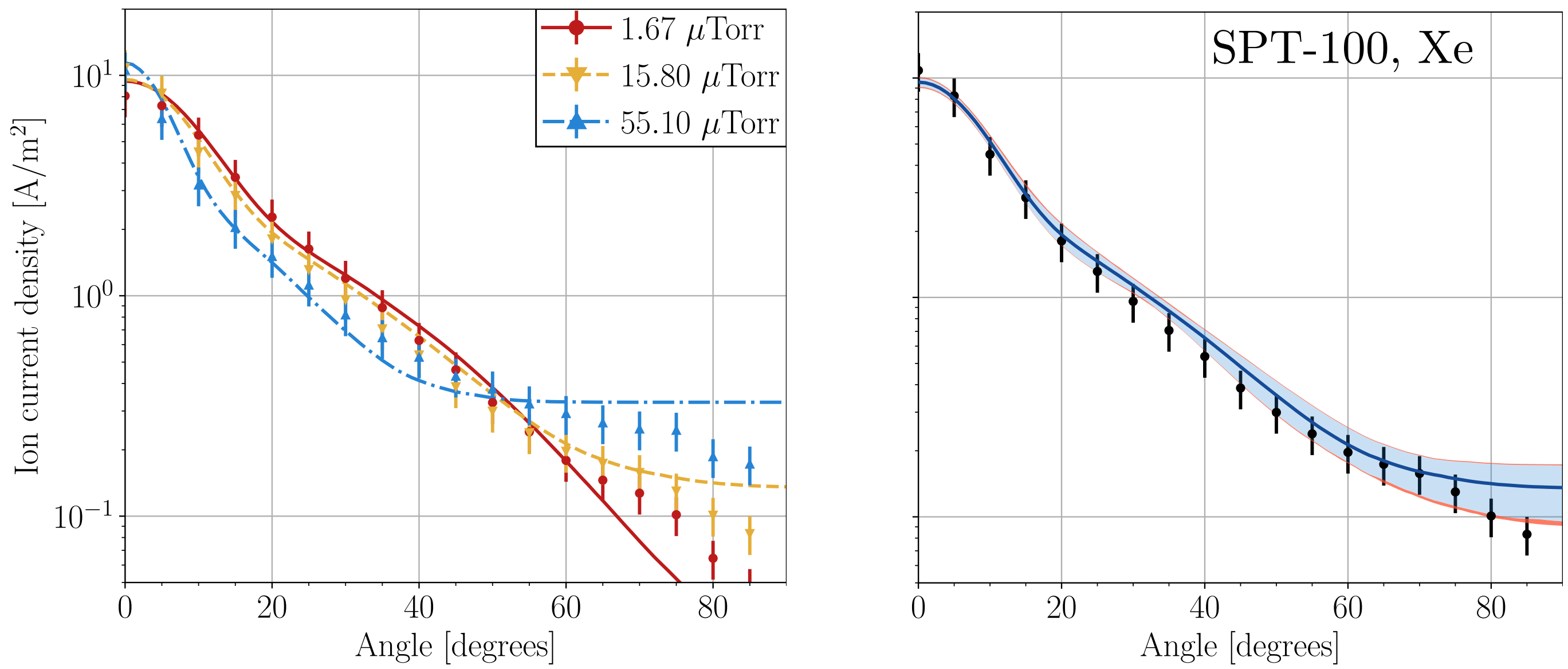

(a) (b)

FIG. 8: (a) Median of posterior predictions of the SPT-100 plume ion current density distribution at $r = 1\,$m compared to data from Ref.[20]. (b) Median posterior prediction and uncertainty in ion current density at $r = 1\,$m for the $P_B = 15.80\,\mu$Torr condition.

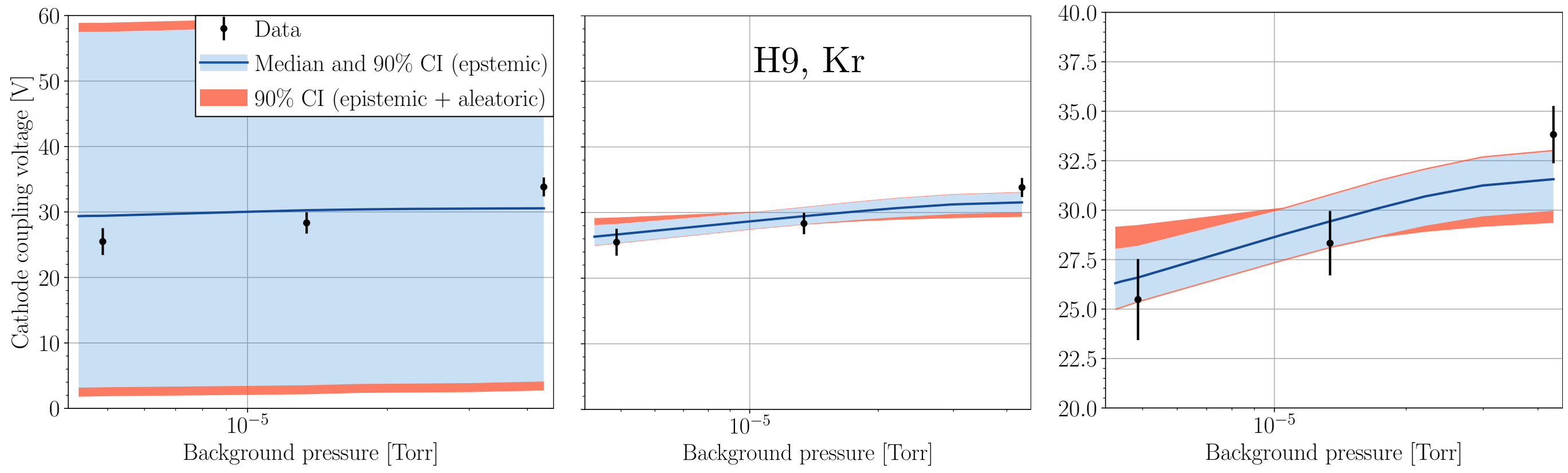

(a) (b) (c)

FIG. 9: (a) Prior, (b) posterior, and (c) zoomed-in posterior predictions of the H9's cathode coupling voltage as a function of background pressure compared to data from Ref 22.

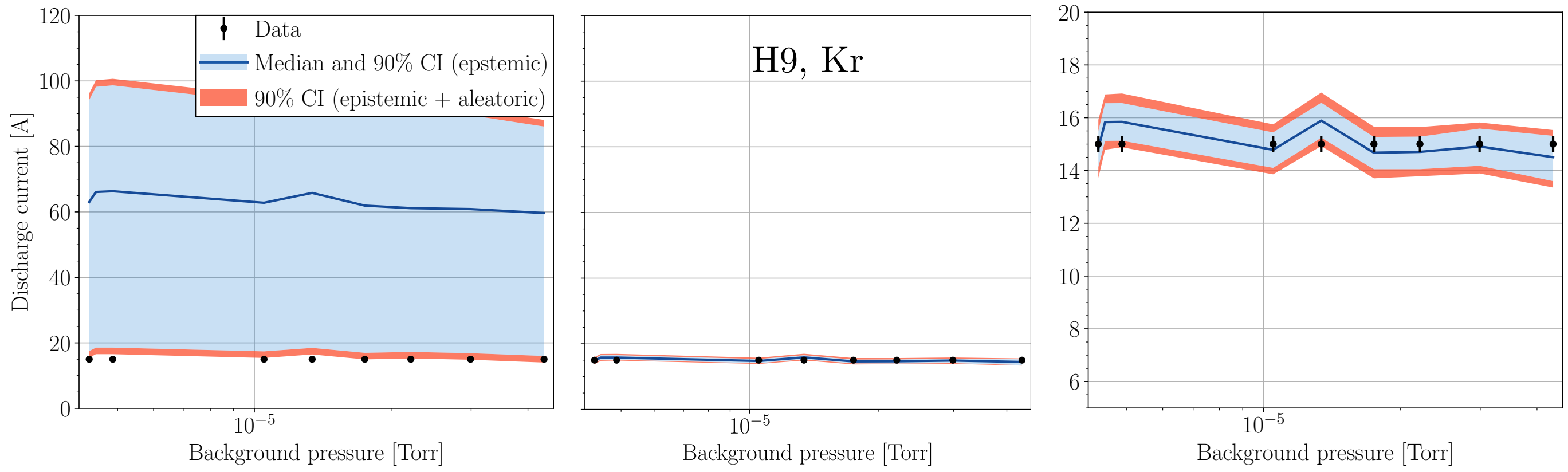

(a) (b) (c)

FIG. 10: (a) Prior, (b) posterior, and (c) zoomed-in posterior predictions of the H9's discharge current as a function of background pressure, compared to experimental data from Ref 22

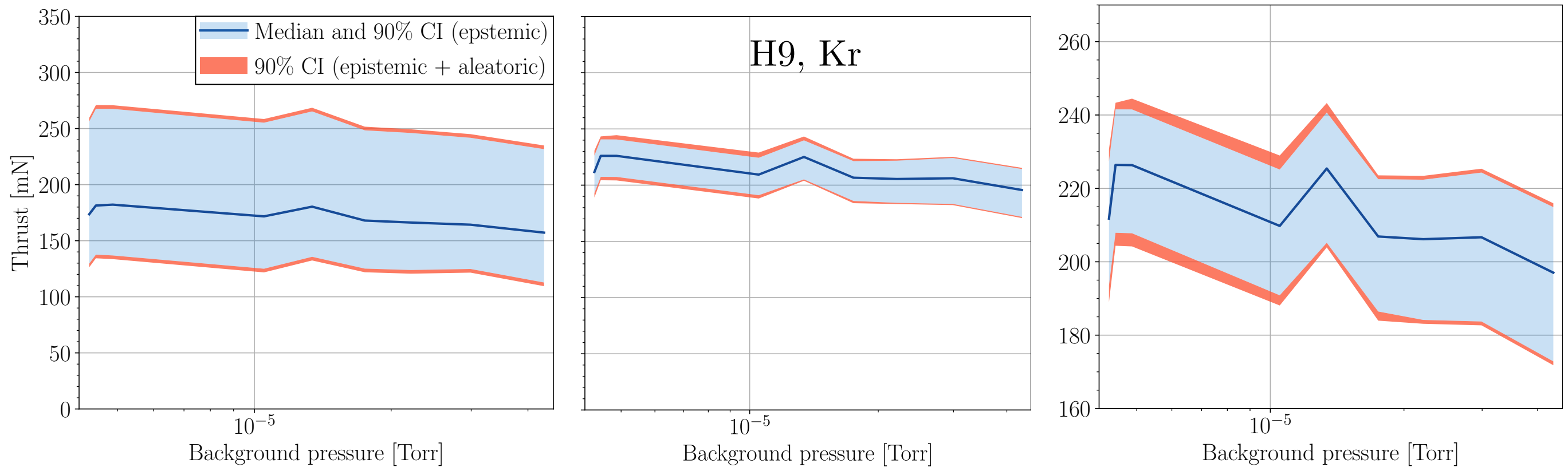

(a) (b) (c)

FIG. 11: (a) Prior, (b) posterior, and (c) zoomed-in posterior predictions of the H9's thrust as a function of background pressure.

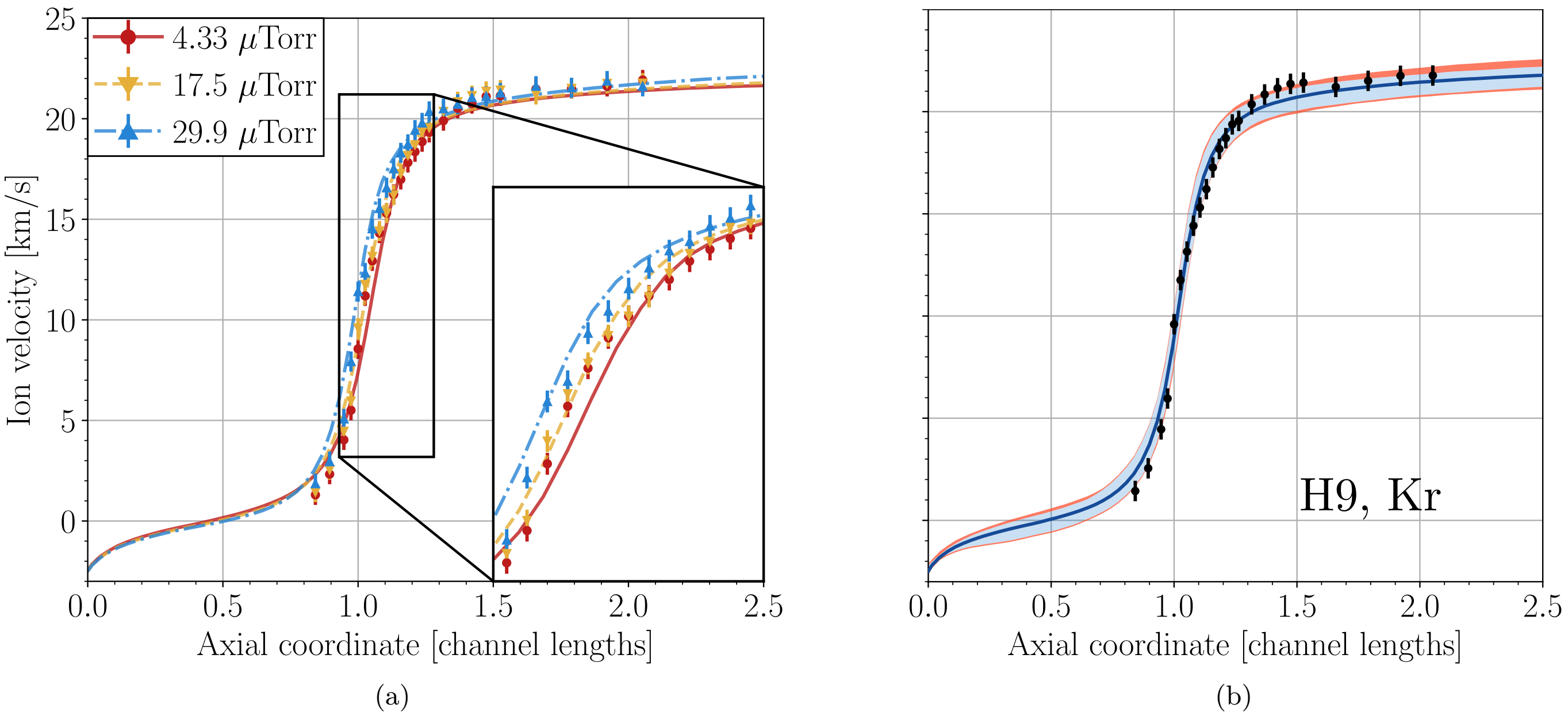

FIG. 12: (a) Median of posterior predictions of the H9's axial ion velocity at three background pressures, compared to data from Ref. 21 (b) Median posterior prediction and uncertainty in ion velocity at the $P_B = 17.5\,\mu\text{Torr}$ condition.

a single pressure. Fig. 13a, shows the ion current density curves at distance of 1.32 meters from the thruster and multiple pressures. As our chosen model form requires that the ion current density peak at zero degrees and decay monotonically with increasing angle, we are unable to capture the observed peak in the data at 7 degrees off-axis. However, we successfully reproduce both the maximum current density and the trends with background pressure for angles up to 40 degrees. As in the SPT-100 data, this departure from the data at larger angles is reflected in Fig. 13b as increased relative uncertainty. Unlike in the SPT-100, the divergence angle of the H9 increases with pressure, which directly reduced the thrust at higher pressures. Without thrust data in the training dataset, this trend was unable to be counteracted by changes in other parameters during the calibration procedure. Finally, we plot in Fig. 14 the current density at all four radii in the training dataset, at a fixed pressure of 26.10 µTorr. This demonstrates that our calibrated model accurately captures trends with distance as well as pressure.

#### 3. Training performance

A more quantitative picture of the training performance can be obtained by examining the relative $L_2$ error of the calibrated model with respect to the training data. Given $N$ samples of parameters $\boldsymbol{\theta}$ and operating conditions $\mathbf{d}$ from the prior or posterior distributions, we calculate the mean and standard deviation of the $L_2$ error in a QoI $q$ as

$$E_q(\mathbf{x}) = \sqrt{\frac{||\mathbf{y}_{eq} - f_q(\mathbf{x})||}{||\mathbf{y}_{eq}||}} \tag{9}$$

$$\mu_q = \frac{1}{N}\sum_{j=1}^{N} E_q(\mathbf{x}_j) \tag{10}$$

$$\sigma_q = \sqrt{\frac{1}{N}\sum_{j=1}^{N} \left[E_q(\mathbf{x}_j) - \mu_q\right]^2}, \tag{11}$$

where $E_q(\mathbf{x})$ is the $L_2$ error in QoI $q$ between the model and data for input parameters $\mathbf{x}$. In Tabs. VI and VII, we report these errors for the SPT-100 and H9, respectively for $N = 1000$ samples of the inputs $\mathbf{x}_j$ drawn from the prior and posterior input distributions as described in Sec. II D. For comparison, we additionally report the errors when the model is evaluated at the median parameter values ($\mu_{50} = E(\mathbf{x}_{50})$) as well as the ratio between the error and the nominal relative measurement uncertainties ($\xi$) used in our previous work.[10] A value of $\mu_{50}/\xi \sim 1$ indicates that the model fits the data within the experimental uncertainty. We note that these metrics are identical to those in our previous work, which facilitates a direct comparison for determining improvement in the models; we include in Tab. VI the model and surrogate errors from the results of the previous work (note that a surrogate was not required in the present work due to optimizations in the thruster code).

All QoIs have a posterior $\mu_{50}/\xi$ of order 1 and show large improvement from the prior. For the SPT-100, most QoIs also show improvement over both the model and surrogate from previous work. The error in $V_{cc}$ is higher than the previous model, which is likely explained by the wide posterior distributions of $T_{ec}$ and $P_T$ as discussed in Sec. III A. We note that the goal of the previous work was to calibrate the true model using the surrogate as a proxy, and we show in Tab. VI that the model in the present work shows considerably greater accuracy compared to the previous model for all QoIs (except $V_{cc}$). There are some cases, such as for thrust, where the surrogate from the previous work performs better than our model, but the ultimate goal is accuracy in the true model, for which the present work demonstrates better performance (e.g. by a factor of eight for thrust). In all cases, the standard deviation of the errors, $\sigma$, is higher than our previous work. While this may seem at first like a negative result, it in fact demonstrates the success of our new likelihood function. In that work, the uncertainty in our predictions was very low, and in many cases the data lay well outside of the uncertainty bounds. Our updated procedure allows for larger predictive uncertainty which more accurately captures the state of our knowledge post-calibration.

The H9 model (Tab. VII) fits the data well for all quantities, although here the interpretation of $\mu_{50}/\xi$ is more complicated. For comparing to the SPT-100 data, we use the same nominal measurement uncertainty values, though we note that the uncertainty in the cathode coupling voltage for the H9 was closer to 2.5%. This would bring $\mu_{50}/\xi$ down to 2.24, which is consistent with the error for the SPT-100 cathode coupling voltage.

### B. Test performance

In this section, we assess the ability of the calibrated model to extrapolate to operating conditions outside of the training dataset. To this end, we use the test datasets described in Sec. II B. In Fig. 15, we plot the prior and posterior predictions of discharge current (Fig. 15a) and thrust (Fig. 15b) for the SPT-100 dataset from Ref. 16. We show in Fig. 16 the same QoIs for the H9 dataset in Ref. 23. These plots compare the predicted QoI to the experimental value, with good agreement indicated by points lying close to the dashed black $y = x$ line. We also report in Tab. VIII the same $L_2$ error metrics as in Sec. III A 3.

The SPT-100 model predicts the test data well and outperforms the model from previous work, though with errors larger than those seen for the training data. The standard deviation in thrust and discharge current errors is reduced to below 2.5%, and the posterior median errors are below 10%. Additionally, the model tends to slightly under-predict the experimental thrust. The model predicts the correct discharge current for the H9 with an error of just 1.3%, with a standard deviation of 1.8%; this good agreement is unsurprising in light of the fact that the discharge current was a constant 15 A in both training and test datasets. Despite lacking thrust in the training data, the H9 model is able to improve on the prior predictions for thrust and obtain a median prediction error of 10%. The error standard deviation has additionally been reduced by a factor of four from the prior, but the predicted thrusts underestimate the experimental values in all cases.

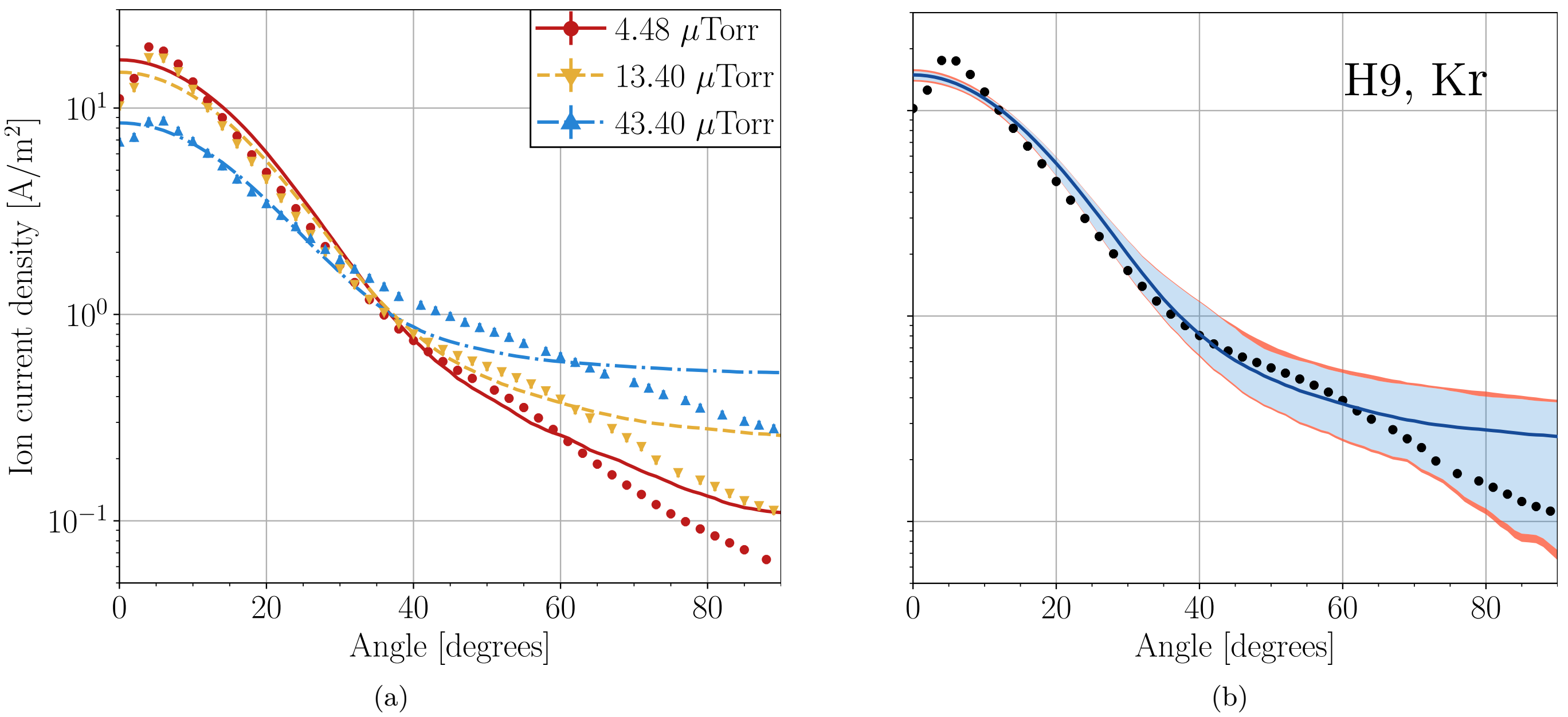

FIG. 13: (a) Median of posterior predictions of the H9 plume ion current density distribution at $r = 1.32\,\mathrm{m}$. (b) Median posterior prediction and uncertainty in ion current density at $r = 1.32\,\mathrm{m}$ and $P_B = 13.4\,\mu\mathrm{Torr}$.

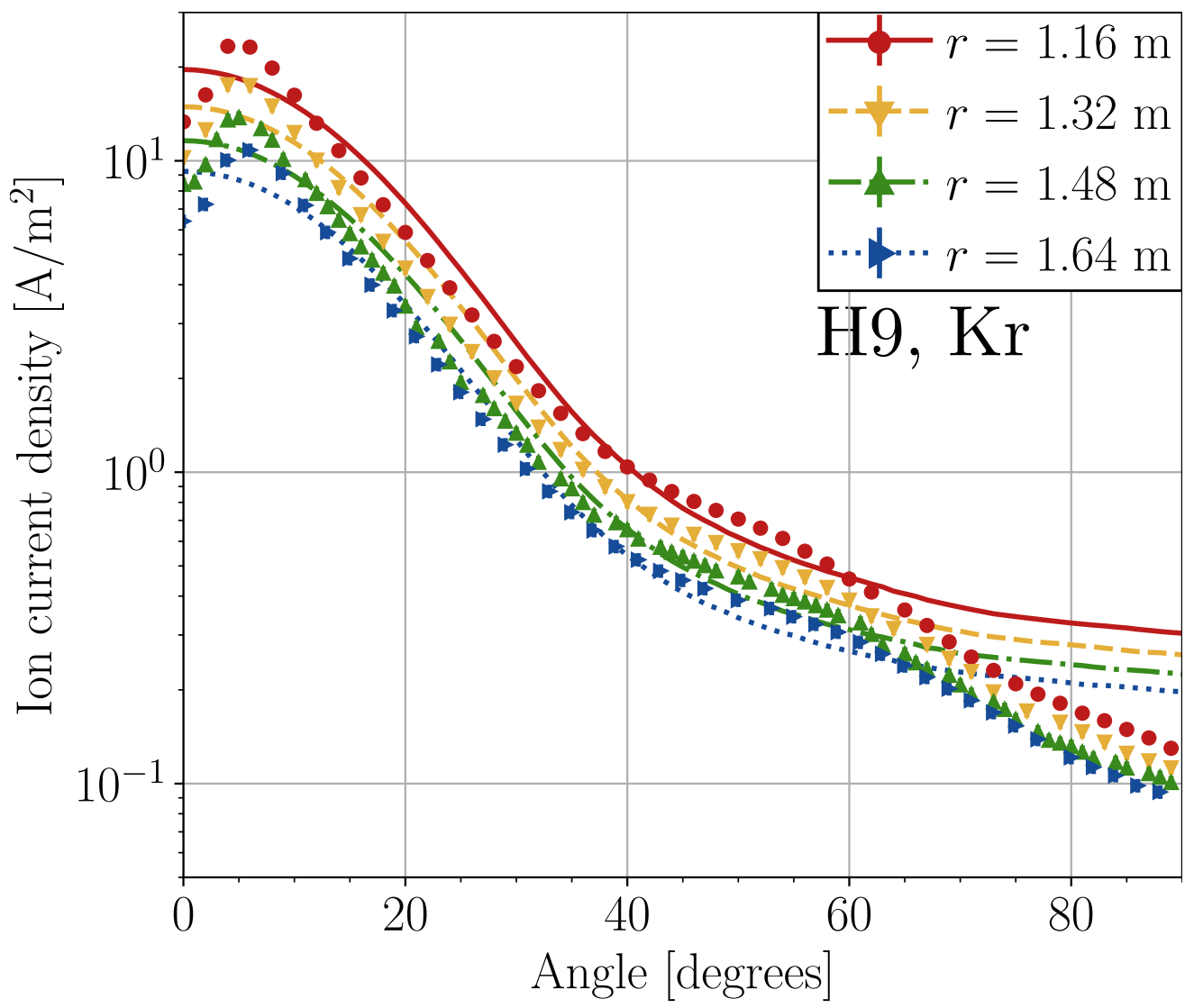

FIG. 14: Median of posterior predictions of the H9 plume ion current density distribution at multiple radii and $P_B = 13.4\,\mu\mathrm{Torr}$.

Taken together, these results show that the calibrated models of both the H9 and SPT-100 are able to extrapolate beyond their training datasets. This was especially observed for the SPT-100, and it is likely that the performance of the H9 model would improve if a wider range of training conditions is made available (i.e., more discharge currents and thrust data).

## C. Extrapolation to orbit

We now turn to using the calibrated model to attempt extrapolation of SPT-100 ground test data to space. We report in Tab. IX predictions of the SPT-100 thrust and discharge current for the on-orbit Express satellite test dataset.[3]

The discharge current is captured to within 5% of the experimental value in both cases, and we correctly predict that the current should increase slightly between the ground and orbit. We also recover the thrust to within 10%, though as in the test dataset from Ref. 16 we underpredict the thrust in both cases. However, we also predict that the thruster should exhibit higher thrust on orbit than on the ground, which conflicts with the trend in the data. We suspect this result stems from the fact that all simulations in the training dataset were performed at a discharge voltage of 300 V, while the SPT-100 from Ref. 3 operated at 310 V on orbit. During training, the model was thus unable to learn how changes in discharge voltage affect the thruster's performance and plasma properties. The effects of the voltage discrepancy between on-ground and on-orbit operation in this case likely overwhelmed the pressure-dependent effects, causing the model to predict the wrong trend. To assess whether the 10 V voltage difference could account for the reversal in the direction of the trend, we run an additional batch of simulations with 300 V on-orbit instead of 310 V (labeled "Orbit (300 V)" in Tab. IX). In this case, we observe a modest reduction in thrust from ground to orbit as expected, with a median decrease of 0.3 mN. We additionally note that the mass flow rate for these data was not measured for either on-ground or on-orbit operation in this dataset, making it challenging to reproduce the operating conditions accurately. These results suggest that more data over a wider range of operating conditions are needed for accurate extrapolation to orbit.

TABLE VI: Relative $L_2$ error between model predictions and training data for the SPT-100. $\xi$ is the nominal experimental error in the data, $\mu$ is the mean error, $\sigma$ is the standard deviation of the error, and $\mu_{50}$ is the prediction error at median parameter values. This work is compared to surrogate and model results from previous work.[10]

| **SPT-100** | | | $L_2$ error [%] | | | |
|---|---|---|---|---|---|---|
| **QoI** | $\xi$ [%] | **Distribution** | $\mu_{50}$ | $\mu$ | $\sigma$ | $\mu_{50}/\xi$ |
| $V_{cc}$ [V] | 1 | Prior (this work) | 4 | 45.9 | 26.8 | 4 |
| | | **Posterior (this work)** | 2.5 | 2.8 | 0.5 | 2.5 |
| | | Posterior (prev. work, model) | 2 | - | - | 2 |
| $T_c$ [mN] | 1 | Prior (this work) | 30.4 | 27.6 | 13.3 | 30.4 |
| | | **Posterior (this work)** | 3.3 | 3.5 | 0.5 | 3.3 |
| | | Posterior (prev. work, model) | 29 | - | - | 29 |
| | | Posterior (prev. work, surrogate) | 2.5 | 2.6 | 0.2 | 2.5 |
| $I_D$ [A] | 10 | Prior (this work) | 728.9 | 667.9 | 324.3 | 72.9 |
| | | **Posterior (this work)** | 3.3 | 3.9 | 1.4 | 0.3 |
| | | Posterior (prev. work, model) | 63 | - | - | 6.3 |
| | | Posterior (prev. work, surrogate) | 45 | 45 | 0.3 | 4.5 |
| $u_{ion}$ [m/s] | 5 | Prior (this work) | 24.2 | 25 | 4.8 | 4.8 |
| | | **Posterior (this work)** | 12.2 | 13.8 | 1.3 | 2.4 |
| | | Posterior (prev. work, model) | 17 | - | - | 3.4 |
| | | Posterior (prev. work, surrogate) | 21 | 21 | 0.2 | 4.2 |
| $j_{ion}$ [A/m$^2$] | 20 | Prior (this work) | 87.2 | 80.7 | 15.4 | 4.4 |
| | | **Posterior (this work)** | 11.4 | 18.6 | 1 | 0.6 |
| | | Posterior (prev. work, model) | 49 | - | - | 2.4 |
| | | Posterior (prev. work, surrogate) | 33 | 33 | 0.3 | 1.6 |

TABLE VII: Relative $L_2$ error between model predictions and training data for the H9. Symbols have the same meanings as Tab. VI.

| **H9** | | | $L_2$ error [%] | | | |
|---|---|---|---|---|---|---|
| **QoI** | $\xi$ [%] | **Distribution** | $\mu_{50}$ | $\mu$ | $\sigma$ | $\mu_{50}/\xi$ |
| $V_{cc}$ [V] | 1 | Prior | 10.7 | 53.8 | 28.4 | 10.7 |
| | | **Posterior** | 5.4 | 6.1 | 1.3 | 5.4 |
| $I_D$ [A] | 10 | Prior | 318.2 | 301.9 | 157 | 31.8 |
| | | **Posterior** | 3.4 | 4.3 | 1.1 | 0.3 |
| $u_{ion}$ [m/s] | 5 | Prior | 45.8 | 43.6 | 8.1 | 9.2 |
| | | **Posterior** | 4.1 | 5.3 | 1 | 0.8 |
| $j_{ion}$ [A/m$^2$] | 20 | Prior | 82.6 | 76.9 | 18.6 | 4.1 |
| | | **Posterior** | 18.9 | 19.5 | 0.5 | 0.9 |

### D. Anomalous electron transport

The magnitude and scaling of the anomalous electron transport are known to have a large impact on Hall thruster model results compared to other parameters. Here, we briefly analyze the calibrated anomalous electron collision frequency profiles and the uncertainty in the five transport parameters. In Fig. 17, we show the profiles of the anomalous electron collision frequencies for the SPT-100 and H9 at three pressures each. As designed, the profile moves upstream at higher pressures. The uncertainty in the axial position of the profile is very low, as reflected by the distributions of parameters $z_{anom}$ and $L_{anom}$ in Tabs. IV and V. The uncertainty in the magnitude of the anomalous transport at the bottom of the Gaussian trough, $\beta_{anom}$ is similarly low. The posterior for this variable lies in the range $(0.95, 1)$ for both thrusters, reduced from a prior range of $(0, 1)$. In contrast, the maximum magnitude of the anomalous transport (governed by $\alpha_{anom}$) has high uncertainty, spanning at least a factor of two for both thrusters. This result is in line with similar observations by Mikellides and Lopez-Ortega,[7,26] as well as those by Hara and Mikellides,[27] which found that the near-anode anomalous collision frequency has a larger effect thrust and ionization oscillations than it does on the ion velocity profile. This also explains the larger uncertainty in $\alpha_{anom}$ for the H9, as without thrust data the near-anode electron transport was not as constrained as for the SPT-100. Finally, the correlation between $L_{anom}$ and $\beta_{anom}$ may have increased the uncertainty in $\alpha_{anom}$ due to their conflicting effects on discharge current, as described in Sec. III A.

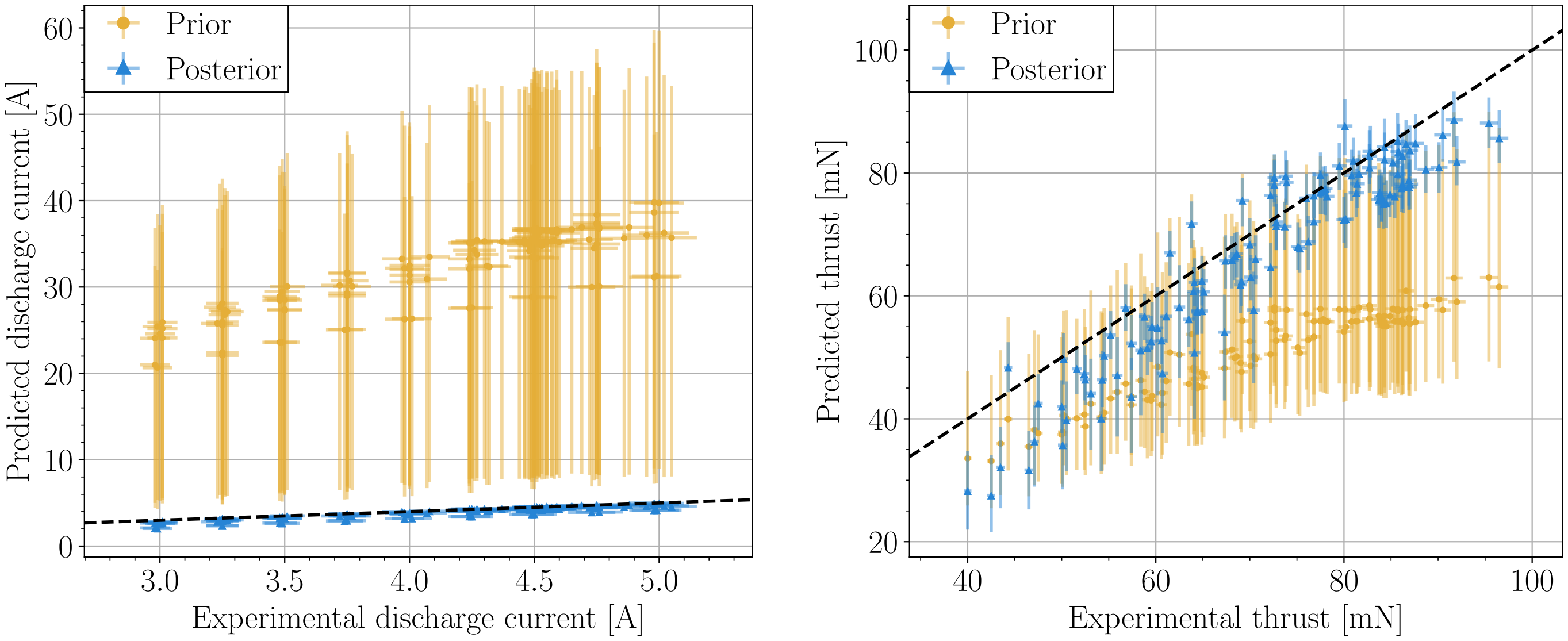

(a) (b)

FIG. 15: Prior and posterior predictions of (a) discharge current and (b) thrust for the SPT-100 Hall thruster from the test dataset in Ref. 16. Perfect agreement ($y = x$) is indicated by the dashed black line. Horizontal error bars indicate the experimental error while vertical bars represent the range between the 5th and 95th percentiles of predictions. The median prediction is indicated by a marker.

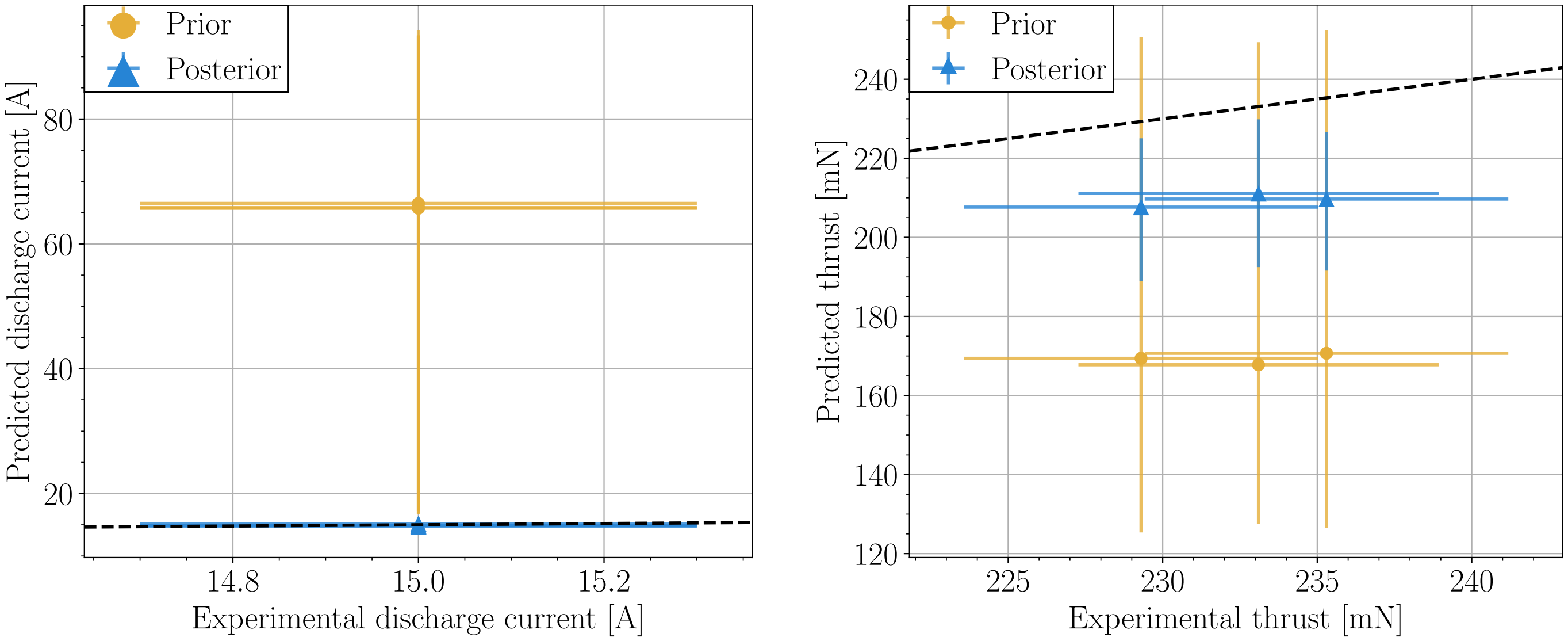

(a) (b)

FIG. 16: Prior and posterior predictions of (a) discharge current and (b) thrust for the H9 Hall thruster from the test dataset in Ref. 23). Plot features have the same meanings as in Fig. 15.

TABLE VIII: Relative $L_2$ error between model predictions and test data for the SPT-100 and H9. The SPT-100 test data comes from Ref. 16 and the H9 test data from Ref. 23. Previous work values come from Ref.[10]. Symbols have the same meanings as Tab. VI.

| **SPT-100** | | | $L_2$ error [%] | | | |
|---|---|---|---|---|---|---|
| **QoI** | $\xi$ [%] | **Distribution** | $\mu_{50}$ | $\mu$ | $\sigma$ | $\mu_{50}/\xi$ |
| $T_c$ [mN] | 1 | Prior (this work) | 29.8 | 28.3 | 12 | 29.8 |
| | | **Posterior (this work)** | 9.6 | 10.4 | 1.8 | 9.6 |
| | | Posterior (prev. work, model) | 30 | - | - | 30 |
| | | Posterior (prev. work, surrogate) | 7 | 7 | 0.1 | 7 |
| $I_D$ [A] | 10 | Prior (this work) | 672.9 | 622.2 | 325.1 | 67.3 |
| | | **Posterior (this work)** | 8.9 | 9.6 | 2.4 | 0.9 |
| | | Posterior (prev. work, model) | 53 | - | - | 5.3 |
| | | Posterior (prev. work, surrogate) | 40 | 40 | 0.1 | 4 |
| **H9** | | | $L_2$ error [%] | | | |
| **QoI** | $\xi$ [%] | **Distribution** | $\mu_{50}$ | $\mu$ | $\sigma$ | $\mu_{50}/\xi$ |
| $T_c$ [mN] | 1 | Prior | 27.2 | 25.6 | 14.8 | 27.2 |
| | | **Posterior** | 10 | 10.4 | 3.8 | 10 |
| $I_D$ [A] | 10 | Prior | 340 | 315.1 | 156.3 | 34 |
| | | **Posterior** | 1.3 | 3.8 | 1.8 | 0.1 |
| $j_{ion}$ [A/m$^2$] | 20 | Prior | 88.6 | 81.3 | 17.3 | 4.4 |
| | | **Posterior** | 34.3 | 34.4 | 1.8 | 1.7 |

TABLE IX: SPT-100 thrust and discharge current from the Express-A satellites[3] compared to model.

| **QoI** | **Case** | **Data** | **Sim. median** | **Sim. 5th pctile** | **Sim. 95th pctile** |
|---|---|---|---|---|---|
| Thrust [mN] | Ground (300 V) | $84.6 \pm 2.4$ | 75.2 | 71.2 | 78.7 |
| | Orbit (310 V) | $83.3 \pm 3.2$ | 76.6 | 73.1 | 80.3 |
| | Orbit (300 V) | - | 74.8 | 70.9 | 78.6 |
| Discharge current [A] | Ground (300 V) | 4.5 | 4.3 | 4.06 | 4.53 |
| | Orbit (310 V) | $4.6 \pm 0.1$ | 4.33 | 4.1 | 4.58 |
| | Orbit (300 V) | - | 4.3 | 4.07 | 4.53 |

### E. Global sensitivity analysis

As in our previous work, we use Sobol's method[28] to compute the influence of each of the model parameters in Tab. I on each of the five quantities of interest in Tab. II. For the ion velocity and ion current density, we use the value attained at the exit plane and peak current density, respectively, as our output variables. We draw 5000 samples from the prior distributions of each variable and use these to estimate the Sobol' total-effect indices. These measure the amount of the variance in each output variable is attributable to any given input,[29,30] including all second- and higher-order correlated effects. Fig. 18 shows the computed indices for each variable and quantity of interest. Note that the sum of all total-effect indices for a given quantity will in general be greater than one.

For both the SPT-100 (Fig. 18a) and the H9 (Fig. 18b), the cathode coupling voltage is most sensitive to the vacuum coupling voltage $V_{vac}$. As this parameter is responsible for setting the minimum coupling voltage, this high level of sensitivity makes sense in light of results in Figs. 4 and 9, in which the change in cathode coupling voltage with background pressure is shown to be small relative to this minimum value. The discharge current is most sensitive to the anomalous transport scale $\alpha_{anom}$ and the transport barrier depth $\beta_{anom}$, with the other parameters having minor effects. The thrust sensitivities are similar for both thrusters, with $\alpha_{anom}$ and multiple plume parameters ($c_0$, $c_1$, and $c_3$) providing the largest contributions, these latter via their influence on the plume divergence efficiency. The ion velocity at the exit plane is most sensitive to the $z_{anom}, \alpha_{anom}, and \beta_{anom}$ anomalous parameters in both thrusters, and the peak ion current density is most influenced by $c_0$, $c_3$, and $c_4$. Lastly, as suggested by the marginal distributions in Sec. III A, all QoIs are insensitive to the neutral ingestion factor $f_n$, with the largest effect seen in the thrust of the SPT-100.

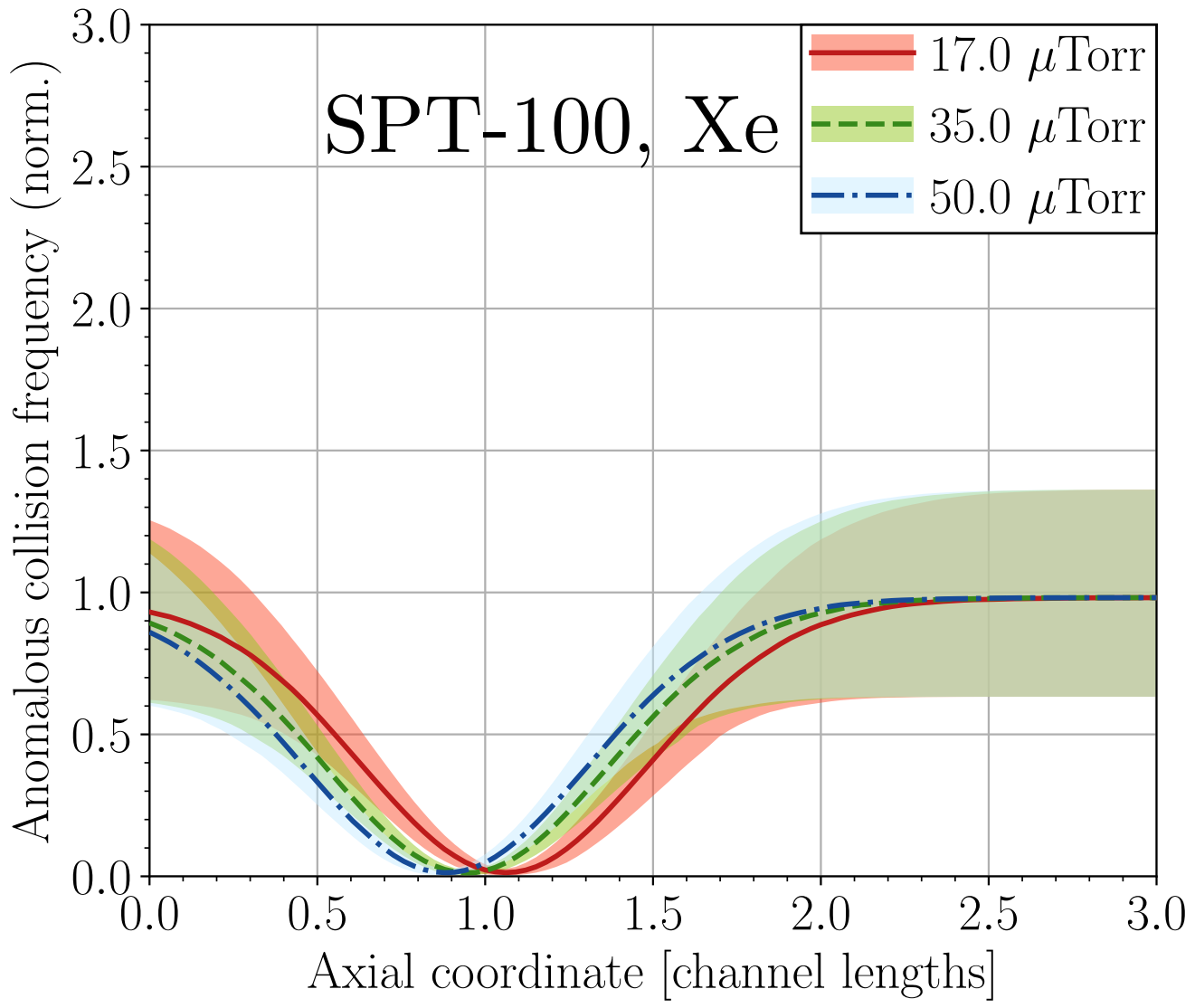

(a)

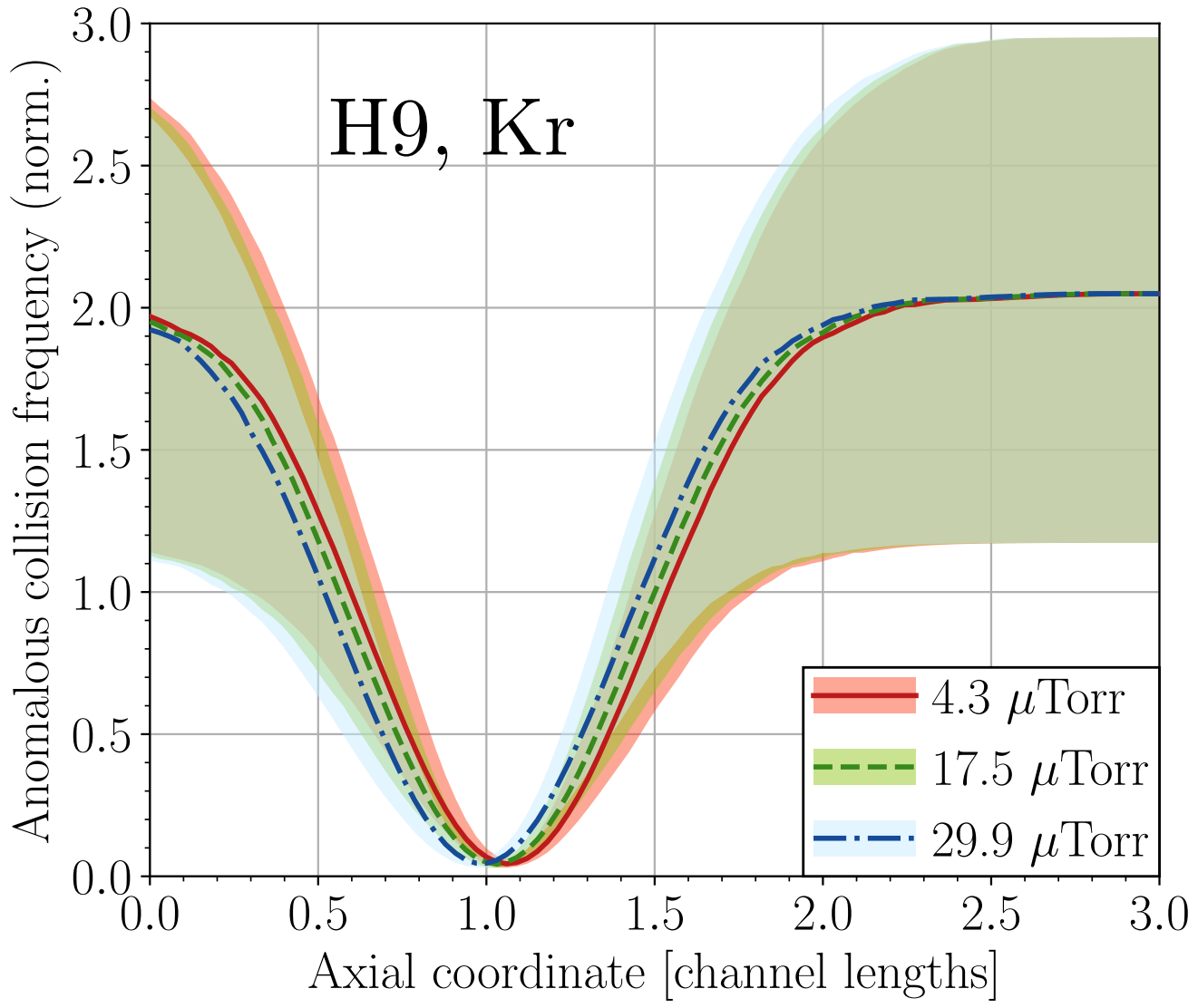

(b)

FIG. 17: Posterior estimates of the anomalous electron collision frequency for the (a) SPT-100 and (b) H9, at three background pressures each. The collision frequencies are normalized by the Bohm diffusion value of $\nu_{bohm} = \omega_{ce}/16$.

The lack of sensitivity of the QoIs to many of the variables, in particular the neutral velocity and the cathode parameters, is likely due to the outsized effects of the anomalous transport parameters on the prior. On restricted parameter ranges closer to the calibrated values, we expect that the relative sensitivity of the QoIs to many parameters would increase. We investigate this hypothesis in Appendix D by repeating the same analysis using samples from the calibrated posterior distributions for each thruster.

## IV. DISCUSSION

In this work, we applied Bayesian inference to calibrate a coupled multi-component Hall thruster model against experimental data for the SPT-100 and H9 thrusters. We used these models to produce probabilistic predictions of several quantities of interest (QoIs), including thrust and spatially-resolved ion velocity, at different operating conditions and background pressures. Across most QoIs, the models of both thrusters exhibited training and test errors of less than 10%, with the SPT-100 model outperforming previous work. We now turn to a discussion of the results, beginning a summary of our core findings. We then discuss some of the challenges we encountered and the primary sources of uncertainty in our predictions. Finally, we dicuss some limitations of our approach and ways in which they might be remedied.

### A. Core findings

1. **Bayesian inference is an effective tool for calibrating and quantifying uncertainty in Hall thruster models.** Our calibration procedure automatically and robustly explored and optimized over a large and high-dimensional parameter space. For all QoIs, it reduced uncertainty in predictions and improved the model's accuracy over the prior without manual intervention. Additionally, we optimized over the entire dataset at once, in parallel, instead of tuning the model parameters per-condition. These results demonstrate the usefulness of Bayesian methods for calibration in the context of Hall thruster modeling.

2. **The calibrated models fit the training data well and can generalize to unseen test data at a limited range of conditions.** The models and parameterizations we used in our coupled framework captured the correct trends in the training data across many background pressures. In addition to high training accuracy across most QoIs, we also observed <10% median test error on thrust and discharge current for both the SPT-100 and H9 thrusters. Additionally, modeling changes in this work related to facility effects and anomalous transport were the primary cause of increased performance over our previous work[10] (see Appendix B). When extrapolating SPT-100 data to orbit, the model correctly captured the trend in discharge current, and obtained median errors of <10% for both thrust and discharge current. However, we predicted the wrong trend with thrust, likely due to a lack of diversity in operating conditions in the training dataset. This underscores the need for varied training data when building predictive models.

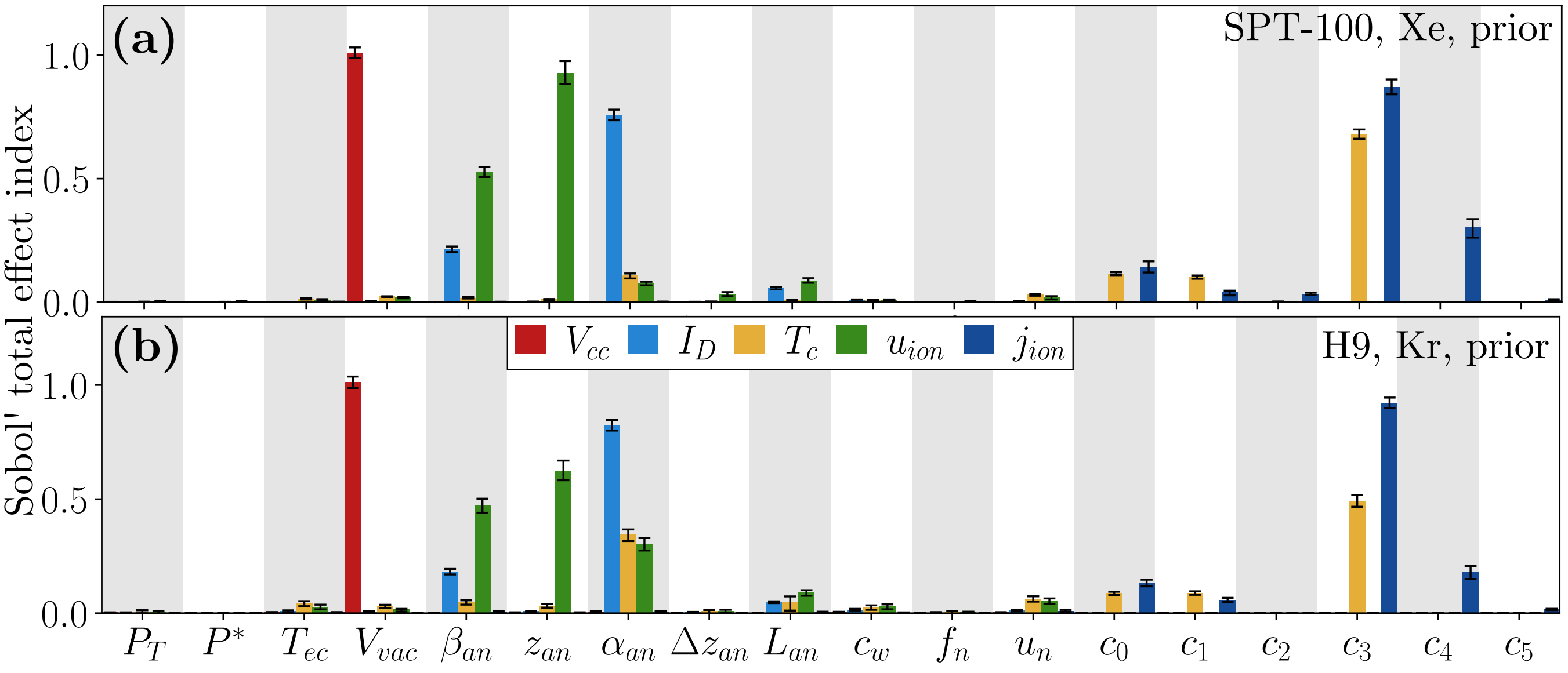

FIG. 18: Sobol' total indices computed over the prior parameter distributions for (a) the SPT-100 operating on xenon and (b) the H9 operating on krypton. Error bars represent 5th and 95th percentile index estimates obtained via bootstrapping with 200 samples.

3. **The phenomenological anomalous transport models were successful at predicting ion velocity profiles at varying pressures.** In this work, we introduced a new four-parameter empirical model for the anomalous electron transport and a simplified logistic pressure-dependent model for the location of the acceleration region. Together, these models were able to reproduce the experimentally-observed trends in ion velocity with high accuracy. In particular, the median ion velocity error for the H9 was less than 5%. Our global sensitivity analysis in Sec. III E showed that, in line with both our previous work[10] and others[6,7,31], the Hall thruster model is highly sensitive to the anomalous electron transport parameters. The four-parameter transport model exhibits a better distribution of sensitivities across its parameters than the two-zone model in our previous work. Additionally, we were able to reduce the number of parameters of the pressure-dependent acceleration region model from four to one without a loss in fidelity. Lastly, we note that neglecting the upstream pressure shift, the posterior anomalous transport parameters were very similar between the H9 and SPT-100. We believe that the four-parameter model with $\alpha_{anom} = 1/16$, $\beta_{anom} \approx 0.99$, $z_{anom} \approx 1.05$, and $L_{anom} \approx 0.38$ may be a good starting choice when simulating other thrusters at 300 V, though it remains necessary to verify this with other codes. However, we note these that parameters may not generalize well to thrusters operating at higher voltages and that an explicit voltage dependence is needed to improve the robustness of the model's extrapolation capabilities.

## B. Aleatoric uncertainty

As noted in Sec. II D, we neglected aleatoric uncertainty in the operating conditions during data. This enabled us to first perform inference on the epistemic variables alone, then propagate forward the aleatoric uncertainty during the prediction step. While convenient, this approach underestimates the true aleatoric uncertainty and breaks down in cases where the aleatoric uncertainty is large. This was the case for the mass flow rate in our on-orbit predictions of the SPT-100, which was only estimated and not experimentally-measured. Accounting for this rigorously would require marginalizing the aleatoric parameters out of the likelihood, potentially using pseudo-marginal MCMC.[24] During prediction, we would then have to sample over a range of operating conditions or provide worst case probabilistic predictions. While this approach would provide a fuller picture of the relative roles of aleatoric uncertainty, it would also incur a larger computational cost. We leave an exploration of these more advanced inference techniques for future work.

### C. Model form error

While we focused primarily on aleatoric and epistemic uncertainty in our work, there remained error due to our modeling and parameter choices. For many QoIs, our chosen model system was capable of capturing the data within epistemic and aleatoric uncertainty, but for others the model and data differed even after calibration. This uncertainty was most evident in predictions of ion velocity for the SPT-100 and ion current density for the H9. In both cases, the epistemic and aleatoric uncertainties were low while large discrepancies from the experimental data remained. This mismatch reflects the inability of the chosen model form to accurately capture features of the data, such as the off-axis peaks in the current density of the H9 or the extended ion backflow region of the SPT-100. The neutral ingestion model was another source of model uncertainty, as it was only partially successful at capturing trends in thrust with background pressure. A robust accounting of model error in a Bayesian context would be a welcome addition to our framework, but is an active area of research.

Without the ability to incorporate model form error into our calibration procedure, we can attempt to reduce it by improving the flexibility of the models themselves. Indeed, much of the improvement from our previous work was enabled by iterative improvements to the thruster component model and its parameterization. The analytic plume model stands our as needing improvement, as it cannot model distributions in which the peak current density occurs away from the thruster centerline, such as in the H9. To remedy this, it would either need to be expanded with additional physical and geometric effects (particularly regarding annular shape of the Hall thruster and the associated focusing of the ion beam), or replaced with a more complex model altogether. Improvements to the cathode coupling model are also possible—in particular, to be able to capture some of the differences between centrally- and externally-mounted cathodes.

The thruster model could also be improved to increase its agreement with data. Charge exchange collisions, not currently modeled by HallThruster.jl, should be a high priority for future versions of the framework, as they play a large role in pressure-related facility effects.[32] Accurately modeling ion acceleration at different discharge voltages and magnetic field strengths would also require a more complex anomalous transport model with explicit or implicit dependencies on these operating conditions.

The sensitivity analysis in Sec. III E shows that the QoIs are sensitive to most of thruster model parameters, with the major exception of the neutral ingestion parameter $f_n$. The insensitivity of the model to this parameter implies the need for a more comprehensive neutral flow model. This model could be analytic, such as that of Frieman et al.,[17] or take the form of a combined model for the plume and vacuum chamber[33]. Incorporating these more complex models requires a much larger amount of information about the thruster and test environment than we currently use, including detailed 2-D thruster geometric and magnetic field information, the distribution of cryopumps and ion gauges throughout the chamber, and the geometry of the facility and beam dump. This information is not always available in the literature, which would pose challenges when calibrating these higher fidelity models.

### D. Choice of quantities of interest

In both this work and our previous work, we compared our simulation results to the same five QoIs—cathode coupling voltage, thrust, discharge current, ion velocity, and ion current density. Each of these QoIs helped the calibration procedure reduce uncertainty for a different set of parameters. Lacking data for one of them, as we did for thrust in the H9 training dataset, led the model's thrust estimates to be poorly-constrained for this thruster. However, the data available for some of our QoIs was limited, potentially reducing their efficacy in driving the calibration procedure toward a predictive mdoel. For instance, at each training point, the discharge voltage was fixed at 300 V for both thrusters, and the discharge current was between 4.25 and 4.5 A for the SPT-100 and exactly 15 A for the H9. This meant that the calibration procedure was not forced to generalize across large changes in mass flow rate, and had no ability to respond to changes in voltage. The impact of this deficiency became apparent in our attempted extrapolation to orbit, in which the uncalibrated voltage response of the model overwhelmed the calibrated pressure-related trends. Including a wider range of conditions in the training data may have reduced posterior uncertainty further and allowed the models to better generalize to the test data.

Additional data sources are available that would further help refine our parameter estimates and motive modeling improvements. For instance, ion energy distribution function measurements made via retarding potential analyzer and species fraction measurements obtained from $E \times B$ probes both provide valuable physical insight into the state and evolution of the Hall thruster plasma. Furthermore, time-resolved data remains an important and underutilized resource, and future versions of our model may be strengthened by attempting to match these data. Non-invasive laser measurements of the electron energy distribution functions[21,34] are an increasingly important source of information about electron transport and plasma heating. In each case, updating our framework to be able to take advantage of these data will require changes to be made to both the component models and the calibration procedure. Finally, carbon back-sputter[35] and electrical and circuit effects[5] stand out as two pressing facility effects which our framework does not attempt to model. Thanks to the modularity and usability changes made for this work, incorporating models for these phenomena would not be prohibitively difficult, and doing so is a priority for future work.

## V. CONCLUSION

In this work, we developed an improved framework for rapid prediction of Hall thruster performance. By coupling cathode, thruster, and plume models with Bayesian inference, we can calibrate model parameters to data and predict important quantities with detailed uncertainty quantification. We used this model to reduce the uncertainty in many key parameters and obtained improved training performance over our previous work. We then demonstrated good generalization outside of the models' training dataset and attempted to extrapolate the performance of a thruster operating on the ground to orbit. Furthermore we extended the model to simulate a magnetically-shielded thruster in addition to the SPT-100. These improvements were made possible due to modeling improvements, including a new empirical anomalous transport model, as well as changes to our Bayesian likelihood.

Despite some challenges, including a lack of thrust data for the H9 and a small range of operating conditions in the training dataset, the current model serves both as a good standalone Hall thruster model as well as a useful baseline for future extensions. By incorporating new and upgraded models as well as additional data sources, we plan to continue improving the framework's predictive power and generality.

## ACKNOWLEDGMENTS

Funding for this work was provided by NASA in part through the Joint Advanced Propulsion Institute (JANUS), a NASA Space Technology Research Institute, under grant number 80NSSC21K1118, as well as in part through a NASA Space Technology Graduate Research Opportunity grant 80NSSC23K1181. This research was additionally supported in part through computational resources provided by Advanced Research Computing at the University of Michigan.

## DATA AVAILABILITY

The code used for this study is open-source and available at `https://github.com/JANUS-Institute/HallThrusterPEM.git`. It builds on the open-source *AMISC* multidisciplinary modeling framework, available online at `https://github.com/eckelsjd/amisc.git`. The Hall thruster component of the model, *HallThruster.jl*, is on GitHub at `https://github.com/UM-PEPL/HallThruster.jl.git`. The datasets generated and analyzed during this study are available from the corresponding author on reasonable request.

### Appendix A: Selected data from the Express satellites

In this appendix, we summarize the data from the Russian Express satellites that we used in our attempt to predict the on-orbit performance of the SPT-100 in Sec. III C. This data was originally reported by Manzella et al. in 2001.[3] Following the analysis of Byrne and Jorns,[36] we use measurements only from thrusters that had been fired for over 30 hours in space, which was 50% longer than the manufacturer's recommended burn-in time. The thrusters meeting this criteria were thrusters RT2 from Express-A #2 and RT1, T4, and RT4 from Express-A #3. We report in Tab. X thrust measurements averaged across these thrusters. We note that the mass flow rate was not measured in either condition and that the discharge voltage on-orbit was 310 V, 10 V higher than the nominal valus from the ground tests.

### Appendix B: Intermediate model results

In this appendix, we show selected results for the SPT-100 Hall thruster using an "intermediate model" between the one in our previous paper[10] and the one presented in this work. We employ the inference and calibration procedure developed in Sec. II C but use a model parameterization similar to our previous work. Specifically, we use a two-zone Bohm-like anomalous transport profile, a four-parameter pressure shift model, and no wall loss or neutral ingestion parameters. This exercise allows us to assess how much of the improvement over our previous work stems from better sampling versus improved modeling.

First, in Fig. 19 we show the discharge current and thrust predicted by the model after calibration. In contrast to the model of the present work, the discharge current of the previous model lies above the experimental value for all pressures and has an exaggerated pressure dependence. The thrust also exhibits overly-strong pressure-related trends. In Fig. 20, we show the predicted ion velocity profiles of the previous model as a function of background pressure. The pressure shift model used in our previous work captures the upstream displacement of the acceleration region well but is over-parameterized, requiring four parameters instead of just one as in the present work. Additionally, the new anomalous transport model better captures the shape and steepness of the ion velocity profiles. Finally, in Tab. XI, we compare the training error metrics of the previous model to those of the present model. We find that the present model exhibits lower mean and median errors in thrust and discharge current by a factor of 60-90%. In all, these findings are approximately as good as those obtained in our previous work, although predict more exaggerated trends in discharge current and thrust with pressure than are observed in the data. These results suggest that the updated model parameterization played a larger role in improving our results from our previous paper than the

TABLE X: Operational and performance data obtained on the ground and in orbit from two Express-A satellites. Data aggregated from Ref. 3. Mass flow rates (*) were not directly measured and are instead calculated in Ref. 3 by assuming a total xenon flow rate of 5.3 mg/s and a 7% cathode flow fraction.

| Pressure [Torr] | $\dot{m}_a$ [mg/s] | $V_D$ [V] | $I_D$ [A] | $T$ [mN] | Note |
|---|---|---|---|---|---|
| $2 \times 10^{-6}$ | 4.29* | 300 | 4.5 | $84.6 \pm 2.4$ | Ground tests |
| $2 \times 10^{-8}$ | 4.29* | 310 | $4.6 \pm 0.1$ | $83.3 \pm 3.2$ | On-orbit measurements |

Reported uncertainties are ± 2 standard deviations.

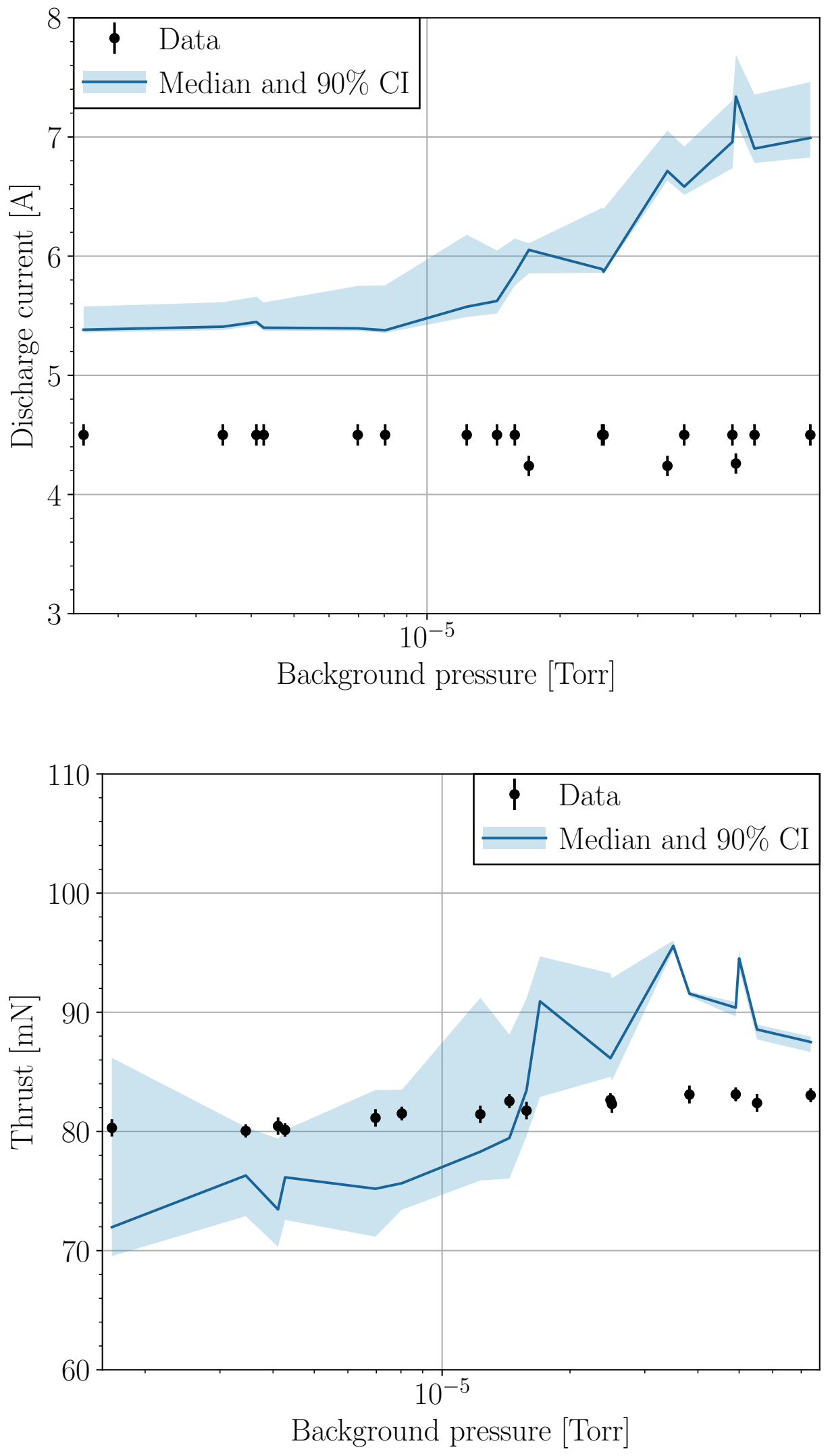

FIG. 19: (a) Discharge current and (b) thrust vs pressure for the SPT-100 using the model of Ref. 10.

changes to the inference procedure.

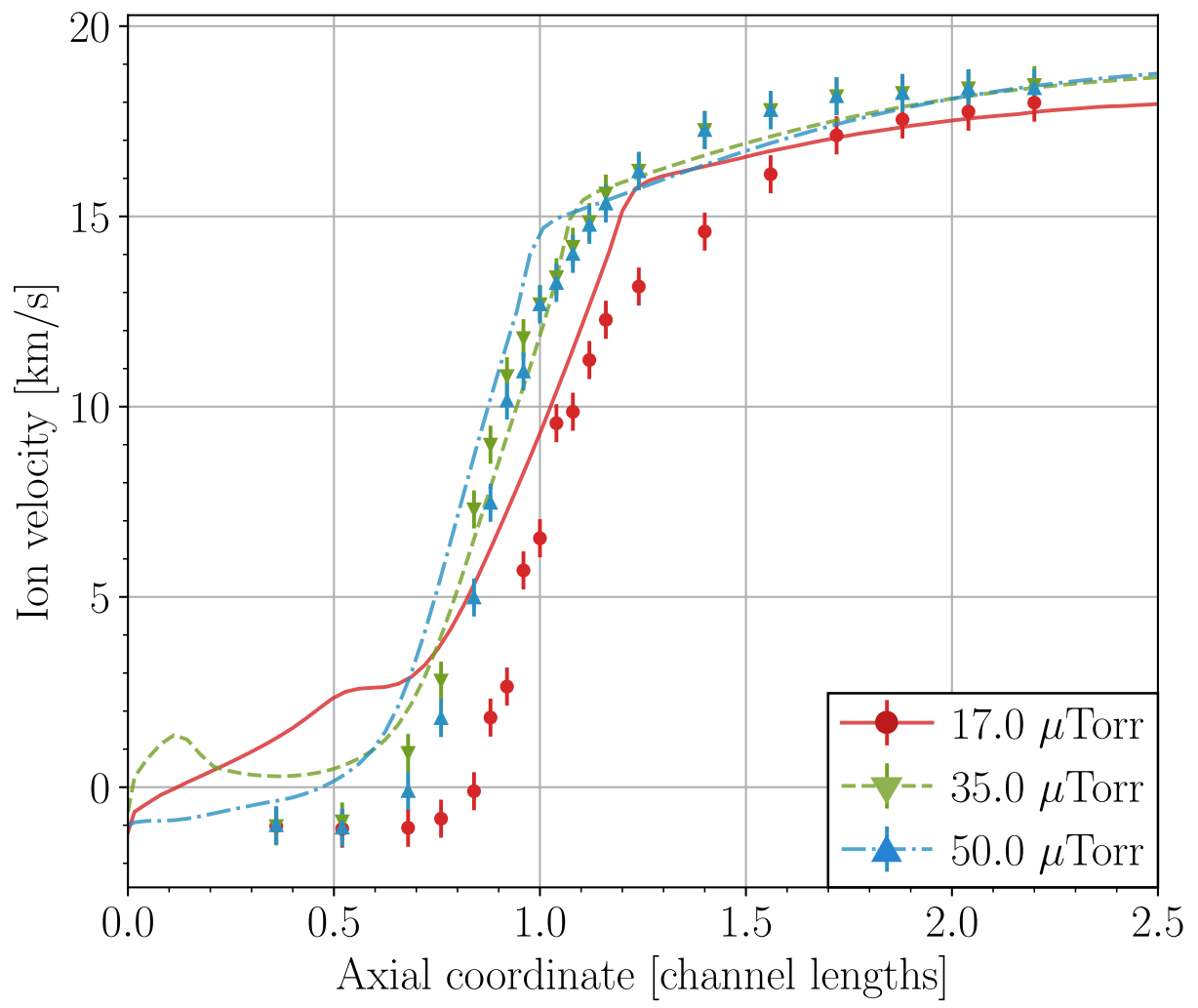

FIG. 20: Ion velocity profiles for the SPT-100 at different background pressures using the model of Ref. 10.

### Appendix C: Posterior parameter distributions

In this appendix, we include plots of the 1-D and 2-D marginal posterior distributions for each model parameter obtained with Bayesian inference. Due to the large number of variables, we have broken these figures up by component for clarity. We plot in Figs. 21, 22, and 23 the cathode, plume, and thruster parameter marginals, respectively, for the SPT-100 operating on xenon. The same distributions for the H9 operating on krypton are in Figs. 24, 25, and 26, respectively.

We make a few observations beyond those made in Sec. III A. For both thrusters, the transport barrier length $L_{anom}$ correlates with $\alpha_{anom}$ and clusters at the upper end of its range, indicating that higher values may have given better results. Additionally, the anomalous pressure shift parameter $\Delta z_{anom}$ correlates with $z_{anom}$, since both parameters shift axially the anomalous electron collision frequency. Across all components, we observe largely unimodal parameter distributions with the exception of $c_2$ for the H9 plume, which has two peaks. This parameter controls the pressure dependence on the divergence angle. Examining Fig. 25, it is not immediately clear why this should be as we fit the trends with pressure well.

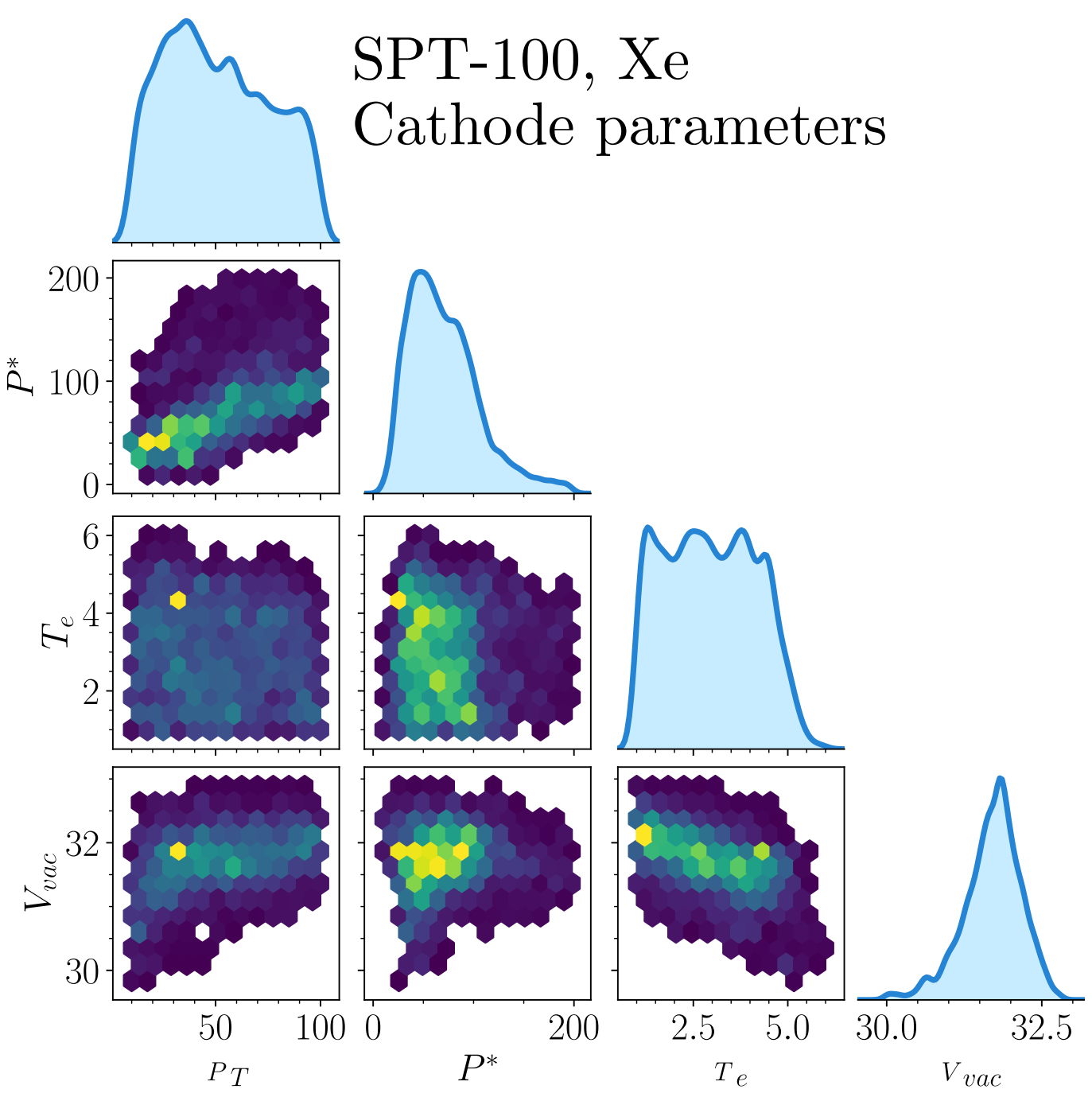

FIG. 21: 1-D and 2-D marginal posterior distributions for the SPT-100 cathode parameters.

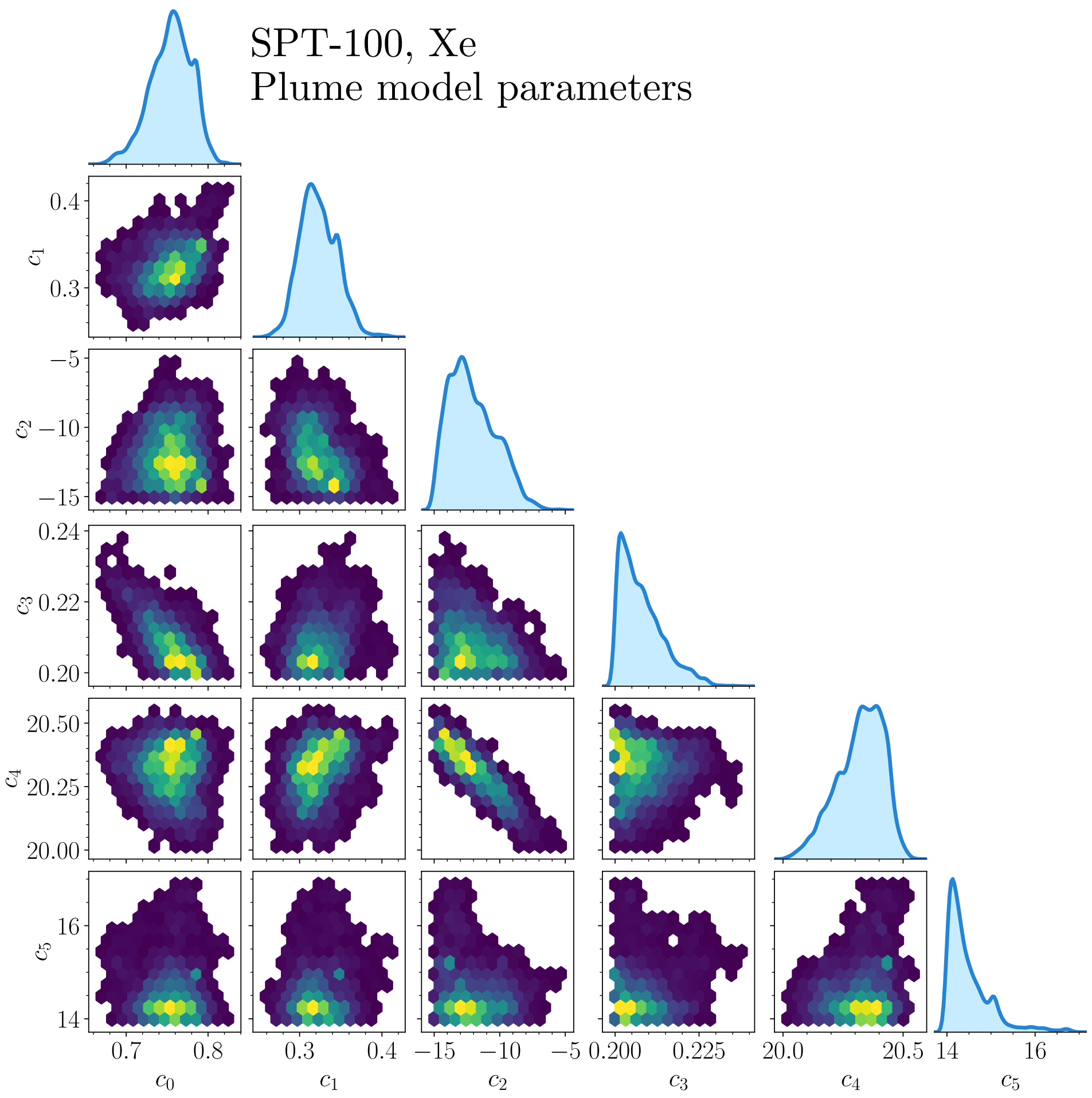

FIG. 22: 1-D and 2-D marginal posterior distributions for the SPT-100 plume parameters.

TABLE XI: Training error metrics of the intermediate model and the model of the present work. Symbols have the same meanings as in Tab. VI

| **QoI** | $\xi$ [%] | **Distribution** | $\mu_{\mathbf{50}}$ | $\mu$ | $\sigma$ | $\mu_{\mathbf{50}}/\xi$ |
|---|---|---|---|---|---|---|
| $V_{cc}$ [V] | 1 | Posterior (old model) | 2.6 | 2.6 | 0.1 | 2.6 |
| | | Posterior (new model) | 2.5 | 2.7 | 0.5 | 2.5 |
| $T_c$ [mN] | 1 | Posterior (old model) | 11.4 | 11.2 | 0.4 | 11.4 |
| | | Posterior (new model) | 3.3 | 3.5 | 0.5 | 3.3 |
| $I_D$ [A] | 10 | Posterior (old model) | 37.6 | 38 | 0.5 | 3.8 |
| | | Posterior (new model) | 3.3 | 3.9 | 1.4 | 0.3 |
| $u_{ion}$ [m/s] | 5 | Posterior (old model) | 13.8 | 16.6 | 0.5 | 2.8 |
| | | Posterior (new model) | 12.2 | 13.8 | 1.2 | 2.4 |
| $j_{ion}$ [A/m$^2$] | 20 | Posterior (old model) | 34.7 | 33 | 0.4 | 1.7 |
| | | Posterior (new model) | 11.4 | 18.7 | 1 | 0.6 |

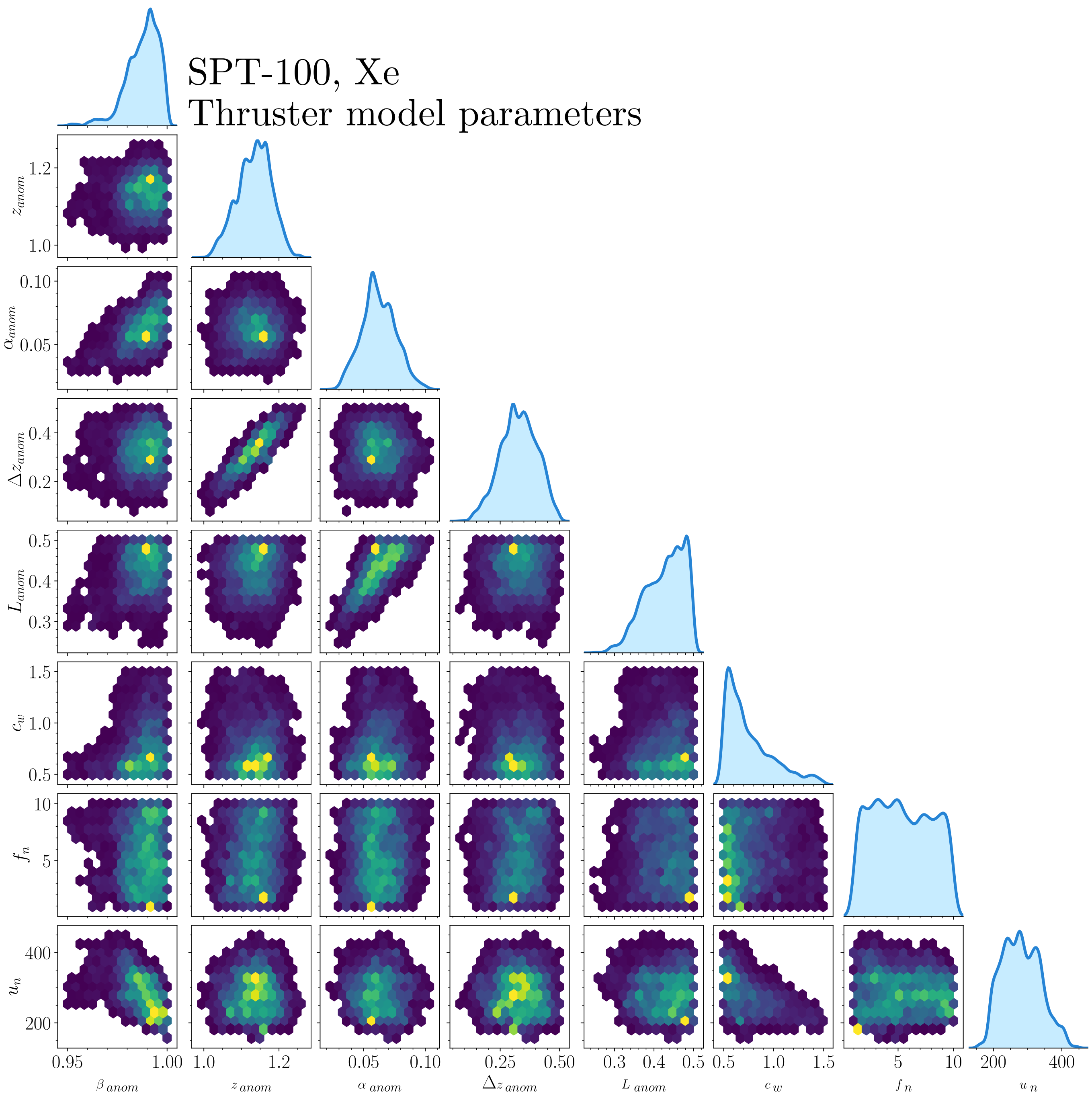

FIG. 23: 1-D and 2-D marginal posterior distributions for the SPT-100 thruster parameters.

#### Appendix D: Sobol' analysis on the posterior

Figures 27a and 27b show the Sobol' total effect indices for the SPT-100 and H9, respectively, computed using samples of the posterior distribution. We caution that these results should not be interpreted quantitatively.[28,30] Variance-based sensitivity analysis assumes the variables are independent and uncorrelated, which is not the case under the posterior. The effects of highly-correlated parameters, such as $z_{anom}$ and $\Delta z_{anom}$, may therefore be mixed and difficult to distinguish from one another. Still, these results are helpful in a qualitative sense to assess the relative importance of the each parameter after calibration, where they have restricted ranges compared to the prior.

Compared to the results obtained from the prior distributions in Sec. III E, the QoIs are sensitive to a larger number of parameters. While the cathode coupling voltage was sensitive mainly to $V_{vac}$ on the prior, on the posterior distributions the influences of the other cathode properties become more important. The SPT-100's more complex cathode coupling voltage curve makes it sensitive to all of the cathode parameters, while the H9 is only sensitive to $V_{cc}$ and $P_T$. Under the posterior, the neutral density and other anomalous transport parameters have also become more important, and for the SPT-100 the thrust is also sensitive to the cathode variables. Lastly, whereas under the prior the peak current density was only sensitive to some of the plume properties, under the posterior it is sensitive to a number of thruster parameters. The more equally-distributed parameter sensitivities under the posterior shows that most parameters have a strong impact on at least one QoI, and that our parameterization is reasonably well-chosen.

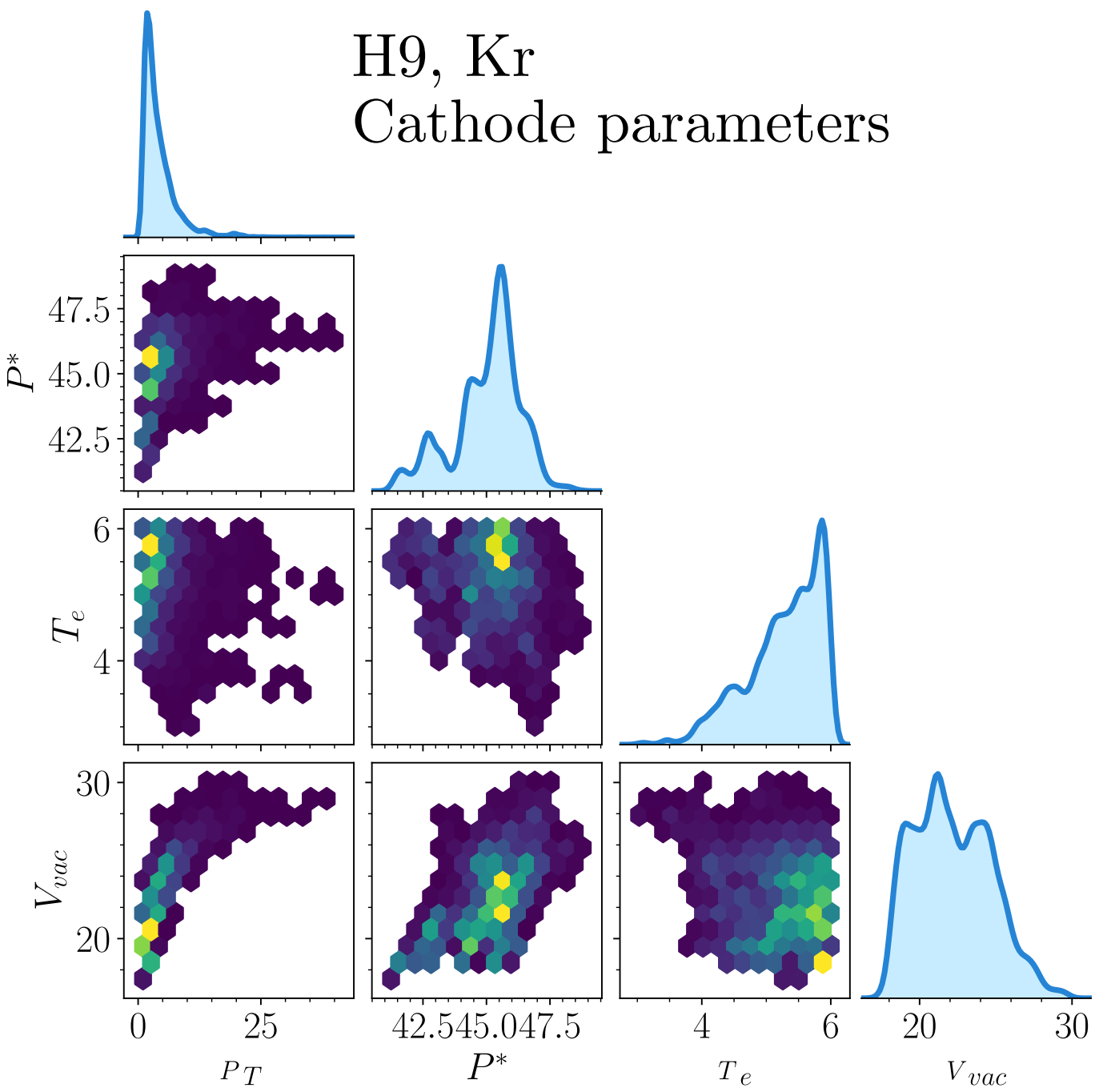

FIG. 24: 1-D and 2-D marginal posterior distributions for the H9 cathode parameters.

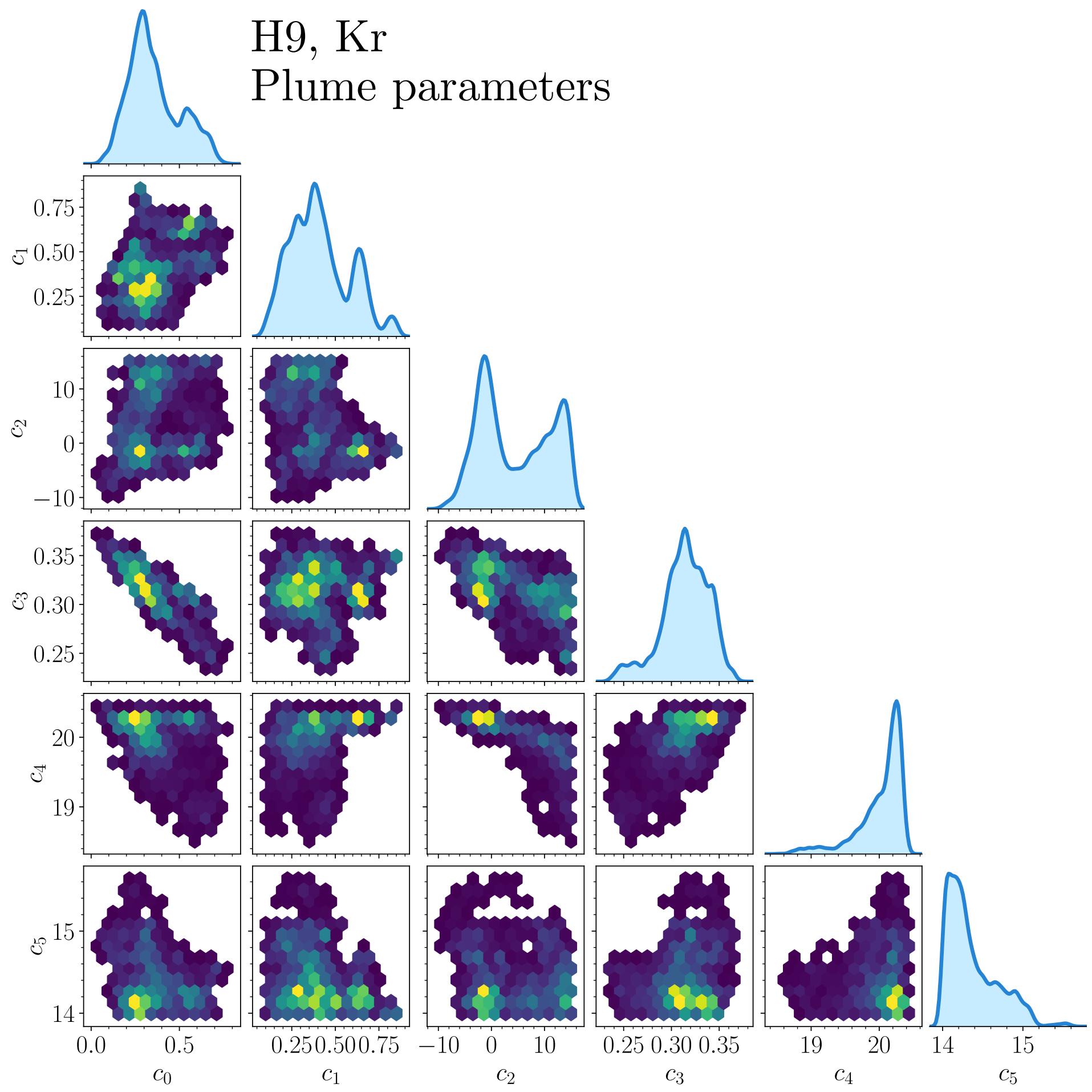

FIG. 25: 1-D and 2-D marginal posterior distributions for the H9 plume parameters.

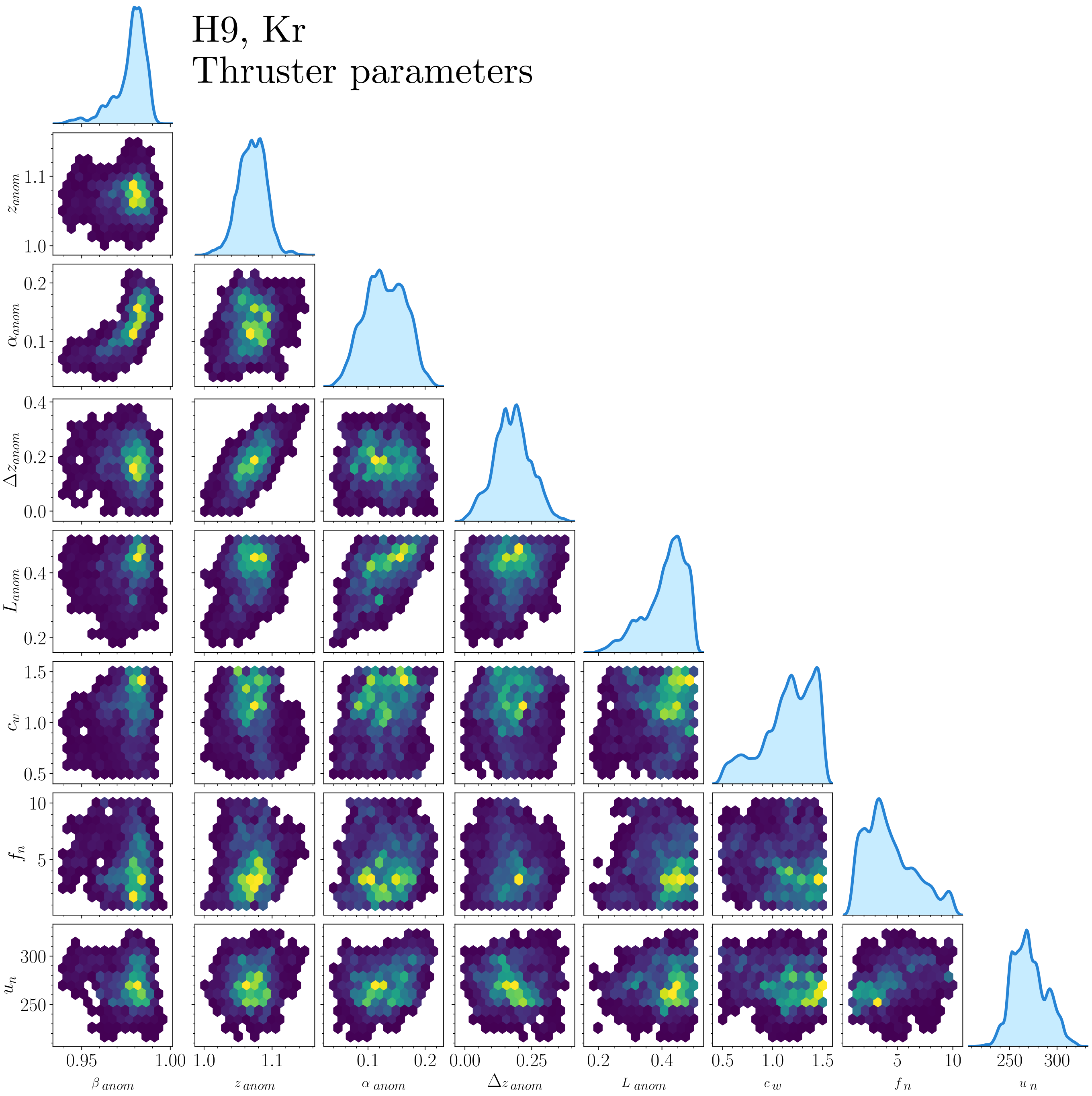

FIG. 26: 1-D and 2-D marginal posterior distributions for the H9 thruster parameters.

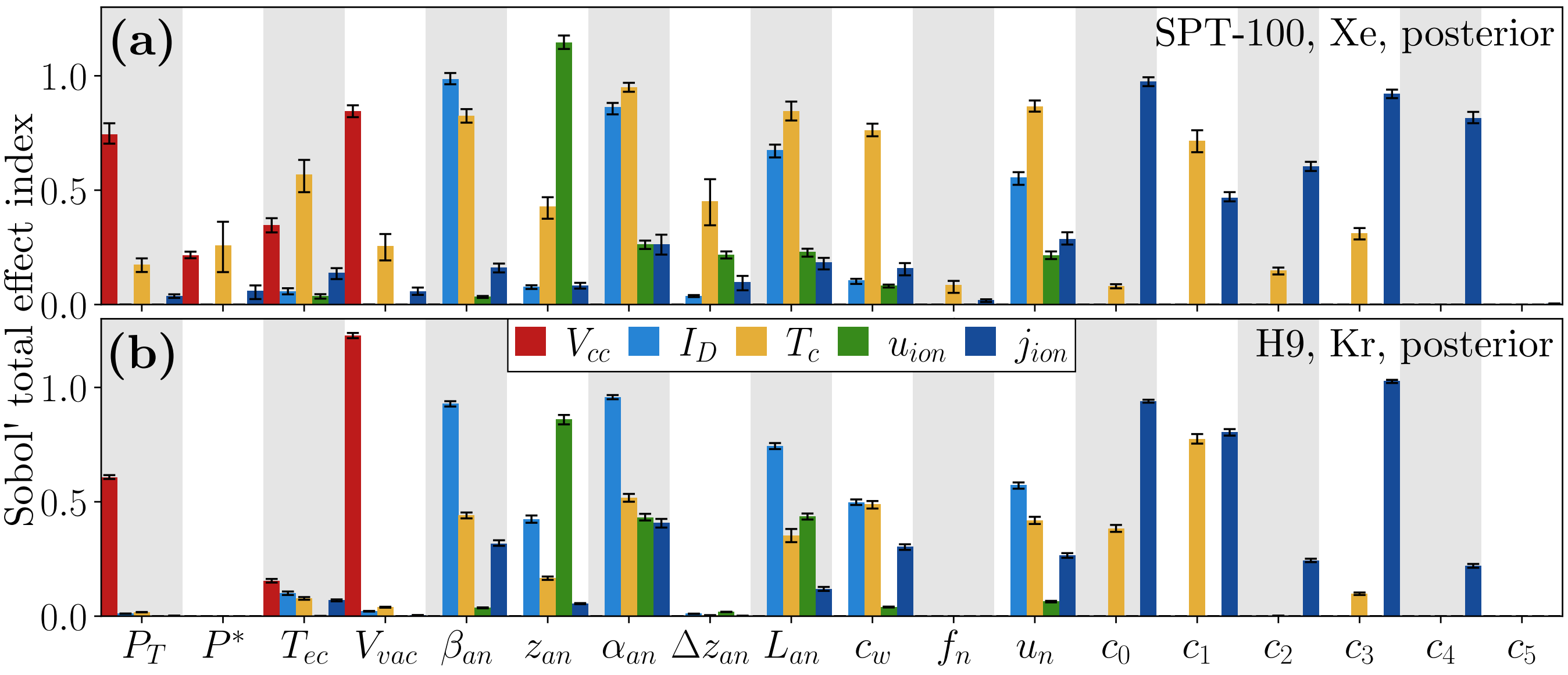

FIG. 27: Sobol' total indices computed over the posterior parameter distributions for (a) the SPT-100 operating on xenon and (b) the H9 operating on krypton. Error bars represent 5th and 95th percentile index estimates obtained via bootstrapping with 200 samples.